\documentclass[journal=jacsat,manuscript=article]{achemso}

\usepackage{caption}
\usepackage{graphicx} 
\usepackage{amsmath}
\usepackage{multirow}
\usepackage{hyperref}
\usepackage{siunitx}
\usepackage{mathtools}
\usepackage[table,xcdraw,dvipsnames]{xcolor}
\usepackage{amsmath}
\usepackage{csquotes}
\usepackage{chemformula}
\usepackage{xr}
\usepackage[T1]{fontenc} 
\usepackage{dsfont}

\DeclarePairedDelimiter{\pqty}{(}{)}
\DeclarePairedDelimiter{\bqty}{[}{]}

\DeclarePairedDelimiterX{\braket}[2]{\langle}{\rangle}{#1\,\delimsize\vert\,\mathopen{}#2}
\DeclarePairedDelimiterX{\mel}[3]{\langle}{\rangle}{#1\,\delimsize\vert\,\mathopen{}#2\,\delimsize\vert\,\mathopen{}#3}

\DeclareSIUnit\atom{atom}
\DeclareSIUnit\angstrom{\text {Å}}
\title{Accelerating first-principles molecular-dynamics thermal conductivity calculations for complex systems}

\author{Sandro Wieser}
\affiliation{Institute of Materials Chemistry, TU Wien, Vienna, Austria}
\author{Yu-Jie Cen}
\affiliation{Institute of Materials Chemistry, TU Wien, Vienna, Austria}
\author{Georg K. H. Madsen}
\affiliation{Institute of Materials Chemistry, TU Wien, Vienna, Austria}
\author{Jesús Carrete}
\email{jcarrete@unizar.es}
\affiliation{Instituto de Nanociencia y Materiales de Aragón, CSIC-Universidad de Zaragoza, Zaragoza, Spain}

\begin{document}

\begin{tocentry}
\includegraphics{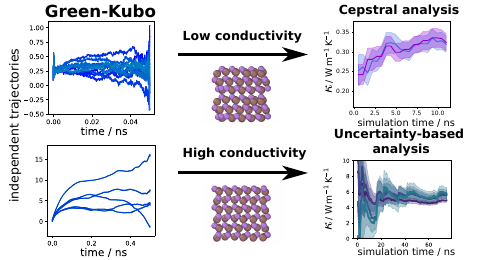}

\end{tocentry}

\begin{abstract}
    
Atomistic simulations of heat transport in complex materials are costly and hard to converge. This has led to the development of several noise-reduction techniques applicable to equilibrium molecular-dynamics (MD) simulations. We analyze the performance of those strategies, taking \ch{InAs} nanowires as our benchmark due to the diverse structures and complex phonon spectra of these quasi-1D systems. We demonstrate how, for low-thermal-conductivity systems, cepstral analysis can reduce computational demands while still delivering accurate results that do not require discarding arbitrary parts of the dataset. However, issues with this approach are revealed when treating high-thermal-conductivity systems, where the thermal conductivity is significantly underestimated. We discuss alternative methods to be used in that situation, relying on uncertainty propagation from independent simulations. We show that the contributions of the covariance matrix have to be included for a quantitative assessment of the error. The combination of these strategies with machine-learning interatomic potentials (MLIPs) provides an accelerated, robust workflow applicable to a diverse set of systems, as our examples using a highly transferable MACE potential illustrate.
\end{abstract}

\section{Introduction}

Heat transport is an integral consideration for a wide range of applications. However, understanding the influence of material-specific aspects on heat transport is quite challenging, in particular for complex systems. In experimental investigations, samples are prone to defects, whose impact usually cannot be measured exactly, making it difficult to obtain concise knowledge transferable to other materials. In atomistic simulations, the defect distribution is well defined, but the calculations tend to be costly or involve significant simplifications; this is particularly problematic if a large number of systems must be studied.

A commonly used family of methods to obtain thermal conductivities is anharmonic lattice dynamics \cite{togo_distributions_2015,Li2014}, which focuses on the direct modeling of heat carriers and their scattering processes. However, this approach is difficult to apply to situations with complex structures or low symmetry, such as defect-laden systems, as well as to scenarios with high anharmonicity. Moreover, there are far fewer implementations of lattice-dynamics workflows targeting lower-dimensional systems, and those available are rarely able to correctly capture the scattering processes. In the technologically relevant case of quasi-1D nanosystems, many publications therefore rely on simple models that extrapolate the vibrational properties of the systems' bulk counterparts and combine them with a very simplified picture of the boundary \cite{li_thermal_2013,karim_temperature-dependence_2020}. However, such approximations fail to properly characterize the physics of the situation, especially for very thin nanowires and for materials where a representative bulk structure is difficult to determine.

Molecular-dynamics approaches offer greater versatility, but non-equilibrium simulations can demand extremely large cells.\cite{Schelling2002, liang_finite-size_2014, Sellan2010} In contrast, equilibrium molecular dynamics with the Green–Kubo (GK) formalism alleviates some of these finite-size issues.
However, achieving proper convergence of the method typically requires long simulation times as well as a system-specific choice of correlation length, which is hard to automate reliably\cite{knoop_ab_2022}. 
A promising method that aims to address these shortcomings is cepstral analysis \cite{ercole_sportran_2022, ercole_accurate_2017}. This approach is based on the denoising of the power spectrum of the heat flux, which is closely related to the GK autocorrelation integral. As the selection of the best result can then be based on objective criteria, the approach allows analysis of uncertainties and has led to a number of successful applications \cite{pegolo_thermal_2024, malosso_viscosity_2022, tisi_thermal_2024}. Another strategy consists in removing parts of the heat flux that do not contribute towards heat transport. For solids this can consist in ignoring the convective transport component \cite{knoop_ab_2022}, while for fluids removing non-diffusive parts can drastically reduce the noise in the resulting correlation functions.\cite{doi:10.1021/acs.jctc.9b01174} Additionally, a method for uncertainty-based analysis, KUTE (green-Kubo Uncertainty-based Transport properties Estimator) \cite{otero-lema_kute_2025}, has recently been developed for GK simulations, but its applicability to thermal conductivity is untested. Yet another technique aims to model the uncertainty of the GK integral using random walks \cite{Oliveira2017}.

While these approaches have been applied successfully in different cases, there is a lack of guidance regarding their respective domains of applicability, how they compare, and which method to apply when facing a new problem. To address this gap, in this work we apply a number of these analysis techniques to the challenging case of quasi-1D nanowires, showcasing both their potential to improve convergence and their system-dependent shortcomings. Specifically, we explore the causes of the poor performance shown by the cepstral analysis technique for certain materials, such as \ch{MgO} \cite{ercole_accurate_2017}. We also investigate different possibilities for uncertainty propagation in the spirit of the KUTE approach to make those uncertainties more quantitative.

To achieve an acceptable computational performance while maintaining a high level of accuracy, we train a MACE model \cite{batatia_mace_2022} on \ch{InAs} bulk, surface, and nanowire structures energies and forces obtained from density-functional theory (DFT). MACE was chosen due to its high level of transferability, demonstrated by recently trained foundation models \cite{batatia_foundation_2023}. A complication with MACE is that its message-passing nature makes the evaluation of the full heat flux through a straightforward implementation extremely expensive. Fortunately, efficient means of obtaining the heat flux for message-passing MLIPs of a different architecture have been derived recently \cite{langer_heat_2023,langer_stress_2023}. We have adapted the original code for use with MACE and made it publicly available \cite{mace-unfolded_2025}.

In this article, we first discuss the creation and validation of the MACE model allowing transferable modeling of thicker \ch{InAs} nanowires than contained in the reference data set, followed by the details of the applied MD methodology. Subsequently, we apply the individual methods to obtain the thermal conductivity of specific target structures and discuss their system-specific benefits and limitations. When such limitations appear, we either suggest alternative system-dependent approaches or provide extensions to the existing methods.

\section{Methods}

\subsection{MACE surrogate models}

MACE \cite{batatia_mace_2022} is a message-passing neural-network force field. MACE models are equivariant with respect to the 3D Euclidean group, meaning that both their internal features and outputs transform consistently as scalars or tensors. For our application we set the maximum order of the spherical harmonics in MACE to $2$, as increasing it to $3$ only led to a marginal increase in accuracy but came at a substantial cost in terms of evaluation speed. The interaction cutoff was set to \SI{5}{\angstrom}, two layers with $64$ uncoupled feature channels were used and the number of message-passing layers was set to $2$. The number of radial Bessel functions was set to 8 and the order of the polynomial employed to achieve the smooth cutoff was set to five. The correlation order of each layer was set to 3 and the width of the three layers of the multilayer perceptron used to build the messages was set to $64$, with SiLU \cite{Hendrycks2016} as the non-linear activation function.

For training, the total data set ---whose composition is discussed below--- was randomly split, with $90\%$ allocated to the training set and $10\%$ to the validation set for each training run. Training the models for $800$ epochs was generally deemed sufficient to achieve proper convergence. For the first $500$ epochs or until no improvement in the loss function was observed for $50$ iterations, the training was performed with a high weight for the forces and a low weight for the energies. For the subsequent $300$ epochs, the energy weight was increased and the force weight reduced. The \textsc{AMSGrad} algorithm\cite{reddi_convergence_2019} was used, with a batch size of $10$.

\subsection{Dataset acquisition and model validation}

\begin{figure}
    \centering
    \includegraphics[width=0.7\linewidth]{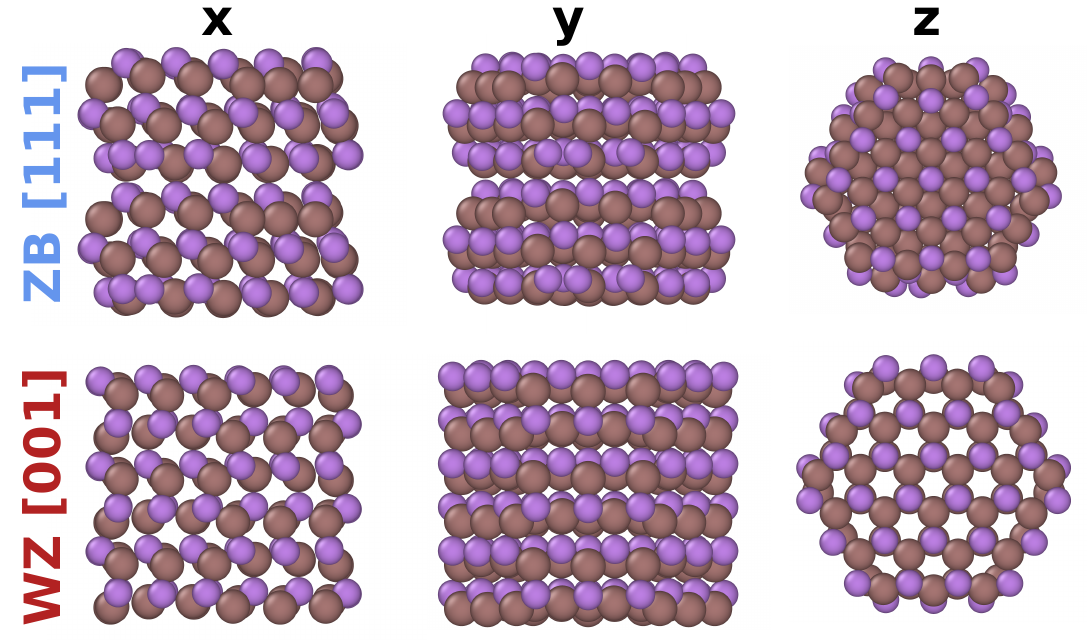}
    \caption{Atomistic structures of the tested zincblende $\bqty*{111}$ and wurtzite $\bqty*{001}$ nanowires from different perspectives.}
    \label{f_atomistic_structures}
\end{figure}

In this work, we focus on heat transport simulations for the two specific \ch{InAs} nanowires shown in Figure~\ref{f_atomistic_structures}. They are hexagonally shaped zincblende (ZB) $\bqty*{111}$ and wurtzite (WZ) $\bqty*{001}$ nanowires. These structures were chosen due to their general similarity at the bulk level and the fact that the WZ nanowire has a much smoother surface than the ZB nanowire, which is likely to cause a large discrepancy of thermal conductivities.

We use a diverse dataset when creating our MLIP, in order to obtain a useful and transferable potential that allows the simulation of a wide range of nanowires beyond DFT-accessible cell sizes, as well as of other \ch{InAs} structures. In fact, we do not include the specific nanowires from Figure~\ref{f_atomistic_structures} directly in the ab-initio dataset, and instead use them for validating the model. Said dataset includes environments corresponding to many different surface configurations, and comprises bulk, surface and nanowire structures. The training was performed using a supervised active learning approach, with four different strategies applied to generate the data: (i) random displacements from starting structures, (ii) random displacements with additional randomly applied biaxial strains from starting structures, (iii) MD simulations with a committee of models followed by a selection of the highest-uncertainty structures, and (iv) optimization of an adversarial loss evaluated from an ensemble of models. The final dataset contains $7357$ diverse DFT-computed structures with up to $180$ atoms per structure. Our dataset is publicly available,\cite{wieser_supplementary_2025} with each structure labeled according to its origin. The following paragraphs give an overview of its composition and more details about the structure generation methods.

As a starting point for the dataset, random displacements drawn from Gaussian distributions with standard deviations ranging from \num{0.01} to \SI{0.1}{\angstrom} were applied to all individual atomic coordinates. For the bulk structures and a selection of the nanowire structures, strains were applied on top of the random displacements. For nanowires, those strains were always parallel to the periodic direction, while for the bulk structure the entire applied strain tensor was randomized according to a normal distribution with a standard deviation of $0.01$. In total, $3115$ different structures were computed based on random displacements.

MD simulations were primarily added as a complementary source of reference data to achieve thermal stability for the structures of interest, as the model quickly diverges into nonphysical territory when trained with just a few initial randomly displaced structures in the database. The procedure was carried out iteratively: The selection from MD simulations was performed employing a Langevin thermostat at a temperature of 300 K in earlier generations of the model and at 700 K once the model reached an improved degree of thermal stability. The simulations were run using a committee of five independent MACE models. Once the MD temperature reached five times the target temperature, the simulation was considered unstable and was aborted. The uncertainty was obtained by aggregating the standard deviations of individual force predictions from the committee in a given configuration using

\begin{equation}
    \label{e_aggregated_force_error}
    \sigma_n = \frac{1}{N_j}\sum_j^{N_j} \frac{1}{3} \sum_a^3 \sigma_{nja} .
\end{equation}

\noindent Here, $j$ denotes the atom index in configuration $n$, $a$ is the index of a Cartesian coordinate, and $\sigma_{nja}$ represents the committee standard deviation of an individual force component.

\begin{figure}
    \centering
    \includegraphics[width=0.5\linewidth]{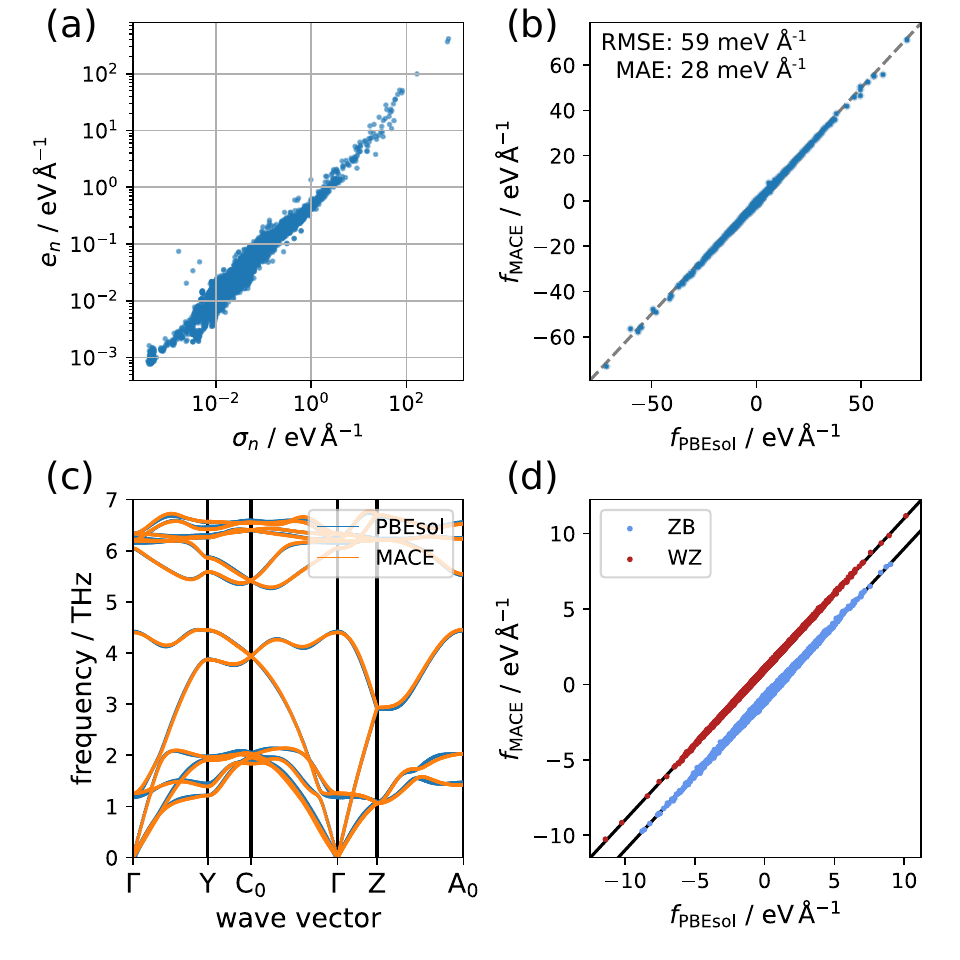}
    \caption{Validation of the approaches for data generation and model training. (a) Correlation between mean force error and aggregated mean force uncertainty on the entire generated dataset, evaluated based on an ensemble of 5 MACE models that were trained on a fraction (4384 configurations) of said dataset. (b) Parity plot for the forces on all structures in the dataset, computed using the final model employed in the production MD simulations. (c) Phonon band structure for the \ch{InAs} WZ phase as evaluated with the final MACE model and DFT. (d) Parity plot for a set of randomly displaced structures for the ZB $\bqty*{111}$ and WZ $\bqty*{001}$ nanowires.}
    \label{f_model_validation}
\end{figure}

In a number of previous works \cite{Carrete2023, heid_spatially_2024} it has been shown that there is a strong correlation between the error and this aggregated force uncertainty, which our own observations confirm (see Figure~\ref{f_model_validation}a). Thus, the structures with the highest uncertainty in their predicted forces were sampled at equidistant intervals to optimize the diversity of the sample. The size of those intervals grew with the number of successfully completed simulation steps and their number never exceeded 10 for each independent MD run. Using this approach, $2145$ unique structures were selected.

Performing long MD simulations to sample a diverse set of reference structures can be inefficient and it could be more meaningful to perform a more conventional maximization of said uncertainty. This can be achieved using an approach based on adversarial attacks as originally employed in \cite{schwalbe-koda_differentiable_2021} and subsequently successfully applied to obtain efficient datasets in various cases \cite{Axelrod2022, Carrete2023, pan_machine_2024, schorghuber_flat_2025}.
The core concept is based on maximizing the logarithmic adversarial loss, formulated as the product of the force uncertainty $\sigma^2$ of an ensemble of models and a Boltzmann factor evaluated at the desired temperature,
\begin{equation}
    \label{e_adv_loss}
    \mathcal{L} = \sigma^2\exp\left({-\frac{E_\mathrm{pot}}{k_\mathrm{B}T}}\right).
\end{equation}
That Boltzmann factor serves to penalize unphysical structures even if they feature a high level of uncertainty. A BFGS algorithm is used in the maximization. Before each optimization, the atomic positions of the initial structures were rattled using a Gaussian distribution with a standard deviation of \SI{0.05}{\angstrom}, with $100$ different random seeds leading to several unique local minima. Based on absolute differences between atomic positions, equivalent structures were discarded and the unique structures appended to the dataset leading to a total of $2097$ unique structures added using this method over the course of several iterations.

Regarding the mixture of bulk phases and nanostructures in the dataset, the starting point was a collection of $500$ bulk ZB and WZ structures generated through randomly applied displacements and biaxial strains ($300$ unstrained and $200$ strained structures). This was deemed enough to accurately describe both bulk phases as judged by the excellent agreement of phonon band structures. Some initial ZB surface configurations were added next, starting with $150$ random displacements. The orientations were chosen as the most likely candidates that would be occurring as facets in the actual nanowires of interest: ZB $\bqty*{0\bar{1}1}$, $\bqty*{112}$ and $\bqty*{111}$ and WZ $\bqty*{100}$. In total, $1332$ different surface slab structures are included in the dataset. 

The largest part of the dataset, comprising a total of $5525$ different structures, is made up of thin nanowire structures that were primarily needed to ensure thermal stability for complex surface environments. The primary component are ZB-based nanowires in $\bqty*{111}$ orientation with hexagonal, circular and square shapes featuring $\bqty*{0\bar{1}1}$ and $\bqty*{112}$ facets. Finally, surface and nanowire structures containing surface defects were also included to further increase the dataset diversity. 

Figure~\ref{f_model_validation}b shows a parity plot based on the entire dataset. The overall agreement is excellent, with a root mean square error (RMSE) of \SI{59}{\milli\electronvolt\per\angstrom} for the forces clearly showing that the model is able to learn the diverse configurations for the different nanowire structures. 

Excellent agreement can also be observed for vibrational properties, obtained using the finite-differences approach as implemented in phonopy \cite{Togo2015}, with a displacement distance of \SI{0.01}{\angstrom}. Figure~\ref{f_model_validation}c shows the comparison of phonon band structures for the WZ bulk phase as evaluated from DFT and the MACE model. The agreement is of similar quality as for the ZB phase, which is shown in Figure~\ref{SI-f_ZB_band_comparison}. This confirms that the addition of non-bulk structures does not degrade the performance on the bulk.

To also showcase the performance for the target nanowires as shown in Figure~\ref{f_atomistic_structures}, in Figure~\ref{f_model_validation}d we compare the force errors on $100$ randomly displaced structures for each nanowire. None of these structures were directly included in the reference data set. Despite this, the RMSE values are in a similar range as those for the training set, with \SI{60}{\milli\electronvolt\per\angstrom} and \SI{36}{\milli\electronvolt\per\angstrom} for the ZB and WZ nanowires respectively. This showcases the transferability of this model to the relevant structures.

\subsection{Ab-initio data generation procedure}

DFT calculations were carried out using the Vienna Ab-initio Simulation Package (VASP) \cite{Kresse1993,Kresse1994,Kresse1996,Kresse1996_2}, using the projector-augmented wave (PAW) method \cite{Kresse1996,blochl_projector_1994}, with the Perdew-Burke-Ernzerhof \cite{Perdew1996,perdew_generalized_1997} exchange and correlation functional revised for solids (PBEsol) \cite{perdew_restoring_2008}, and considering $s$, $p$, and $d$ as valence states for In and only $s$ and $p$ as valence states for As. The plane-wave energy cutoff was set to \SI{450}{\electronvolt} and the Gaussian smearing width to \SI{0.05}{\electronvolt}. For the zincblende structure, the k-point mesh was set to $5\times5\times5$ for the conventional cubic unit cell, and the k-point mesh for all other structures was adjusted to conform to an equivalent density based on the respective lattice parameters, always rounding up. The energy cutoff and k-point density were chosen based on careful convergence tests to achieve an accuracy of \SI{1}{\milli\electronvolt\per\atom}. The energy convergence criterion of the self-consistent procedure was set to \SI{e-8}{\electronvolt} and the convergence for the structure optimizations was set to a maximum force of \SI{e-3}{\electronvolt\per\angstrom}. The projection operators were evaluated in reciprocal space to improve the accuracy.

\subsection{Green-Kubo thermal-conductivity calculations}
Based on the fluctuation-dissipation theorem \cite{Kubo1966, Schelling2002}, the GK method extracts the thermal conductivity from the fluctuations of the heat flux $\mathbf{J}$ during equilibrium MD simulations. Specifically,

\begin{equation}
\label{e_therm_cond}
    \mathbf{\kappa} = \frac{V}{k_B T^2} \int_0^\infty dt \left< \mathbf{J}(t)\otimes\mathbf{J}(0)\right>,
\end{equation}

\noindent where $T$ is the temperature and $V$ is the system volume. The argument of the integral features the heat flux autocorrelation function (HFACF), where the heat flux $\mathbf{J}$ can generally be decomposed into a potential term $\mathbf{J}_{\mathrm{pot}}$ and a convective term $\mathbf{J}_{\mathrm{conv}}$ as \cite{hardy_energy-flux_1968,langer_heat_2023, Hamakawa2025}
\begin{equation}
\label{e_hflux}
    \mathbf{J}(t) = \mathbf{J}_{\mathrm{pot}} + \mathbf{J}_{\mathrm{conv}} = \frac{1}{V}\sum_{i,j} \left[ \mathbf{r}_{ji} \left( \frac{\partial U_i}{\partial \mathbf{r}_j} \cdot \mathbf{v}_j \right) \right] + \frac{1}{V}\sum_i E_i \mathbf{v}_i .
\end{equation}

\noindent Here, $U_i$ and $E_i$ are the potential and total energy of each atom, $\mathbf{r}_{ji}$ the vector connecting the positions of atoms $i$ and $j$, and $\mathbf{r}_j$ and $\mathbf{v}_j$ are the position and the velocity of $j$, respectively. 

For solids with negligible mass transport (such as ionic diffusion) it is reasonable to exclude the convective component of the heat flux in equation~\eqref{e_hflux} to substantially reduce the noise in the HFACF while not affecting the resulting thermal conductivity \cite{Nagoya2022, Zhang2023}. It should be stressed that this is an approximation also for crystals and does lead to minor differences compared to the consideration of the full heat flux. A more rigorous approach would involve the consideration of bounded displacements with respect to equilibrium positions \cite{langer_heat_2023}. However, we will focus on an even more generally applicable approach to reduce the noise, which consists in exploiting the gauge invariance of the heat flux \cite{ercole_gauge_2016} to remove the parts not contributing to the thermal conductivity. This is done by subtracting the contribution from the time average of the atomic virial $S_i$ from the total heat flux \cite{knoop_ab_2022}.

\begin{equation}
    \label{e_gauge}
    \mathbf{J}_\mathrm{gf}(t) = \mathbf{J}_\mathrm{raw} - \frac{1}{V}\sum_i\left< S_i \right>_t\mathbf{v}_i.
\end{equation}

Equation~\eqref{e_hflux} is straightforward to compute for legacy pair potentials.  In contrast, difficulties can arise for many-body potentials due to the lack of a univocal scheme to apportion parts of the total potential energy to each atom. Many modern MLIPs are actually well suited for this task, as they model the total system energy as a sum of atomic terms. While these atomic pseudoenergies carry no specific physical meaning, it has been shown in the context of the aforementioned gauge invariance that the thermal transport coefficients are independent of the partition of the energies among atoms \cite{ercole_gauge_2016}. However, for message-passing MLIPs, where atoms exchange information with atoms beyond their immediate neighborhood as defined by a cutoff radius, this becomes computationally much more involved. This problem was previously addressed introducing a computationally efficient solution based on automatic differentiation to compute the potential term.\cite{langer_heat_2023, langer_stress_2023} This solution requires the expansion of the simulation cell to directly include all atoms within an effective cutoff including all message-passing layers. The potential term of the heat flux is then reformed to 

\begin{equation}
\label{e_hflux_message_passing}
    \mathbf{J}_{\mathrm{pot}} = \frac{1}{V}\sum_{j \in \mathcal{R}_{\mathrm{unf}}} \frac{\partial\mathbf{B}}{\mathbf{\partial \mathbf{r}}_j} \cdot \mathbf{v}_j - \frac{1}{V}\sum_{j \in \mathcal{R}_{\mathrm{unf}}} \mathbf{r}_j \left( \frac{\partial U}{\partial \mathbf{r}_j} \cdot \mathbf{v}_j \right).
\end{equation}

\noindent Here, the index $j$ runs over all atoms in the simulation cell including the atoms that must be considered when the cell is \enquote{unfolded} (extended) up to the effective cutoff radius. In the derivative of the \emph{potential barycenter} $\mathbf{B}=\sum_{i\in \mathcal{R}_{\mathrm{cell}}}\mathbf{r}_i U_i$, where the index $i$ runs over all atoms in the actual cell, $\mathbf{r}_i$ is excluded from the graph used for automatic differentiation and therefore does not affect the derivative. This is done to speed up the computation, as the derivative of a scalar with respect to the position vectors, such as $\partial U/\partial \mathbf{r}_j$, is much cheaper to evaluate. For the same reason, in the second term the sum over the atomic potential energies $U_i$ was already included inside the derivative. Using equation~\eqref{e_hflux_message_passing}, the heat flux is computed for the actual atoms in the simulation cell based on all the atoms in the \enquote{unfolded} cell, which is treated in a non-periodic picture. 

Note that for quasi-1D subperiodic systems the lattice thermal conductivity can be treated as a 1D scalar and only the projection of $\mathbf{J}$ along the periodic axis is relevant to the calculation. The same applies to all derived fluxes discussed in this section. Likewise, the outer product in Eq.~\eqref{e_therm_cond} is reduced to an ordinary product between that projection evaluated at two different times. Thus, the additional computational effort required by the \enquote{unfolding} is quite mild, given that only one component of the flux is calculated and that the number of additional atoms when there is only one periodic direction is limited.

We adapted the implementation of the heat flux created in ref.~\citenum{langer_heat_2023} to MACE models, and we validated our implementation with the direct evaluation of equation~\eqref{e_hflux} \cite{mace-unfolded_2025}. 
To perform gauge fixing based on the potential flux for MACE, we reuse the quantities computed for the heat flux:

\begin{equation}
    \label{e_atomic_virial_mace}
    S_j = \begin{cases}
    \frac{\partial\mathbf{B}}{\mathbf{\partial \mathbf{r}}_j} -  \mathbf{r}_j  \otimes\frac{\partial U}{\partial \mathbf{r}_j} + \mathds{1}E_j & j \in \mathcal{R}_{\mathrm{sc}} \\
    \frac{\partial\mathbf{B}}{\mathbf{\partial \mathbf{r}}_j} -  \mathbf{r}_j  \otimes\frac{\partial U}{\partial \mathbf{r}_j} & j \not\in \mathcal{R}_{\mathrm{sc}},
    \end{cases}
\end{equation}

\noindent where the index $j$ runs over the atoms in the unfolded cell and the atomic energy $E_j$ is added to the diagonal elements to include the contribution of the convective heat flux, which is only added to the atoms in the actual simulation cell $\mathcal{R}_{\mathrm{sc}}$ for consistency with equation \eqref{e_hflux}.

The MD runs for all GK simulations were conducted using the Large-scale Atomic/Molecular Massively Parallel Simulator (LAMMPS) \cite{Thompson2022} while the MD simulations used to build the dataset were performed via the integrators available within the Atomic Simulation Environment (ASE) \cite{hjorth_larsen_atomic_2017}. In every case, an integration time step of \SI{1}{\femto\second} was used and an equilibration period of at least \SI{20}{\pico\second} in an NVT ensemble preceded each production run. In a further preparation step right after the equilibration period, the angular momentum of the nanowires was forced to zero by scaling the atomic forces while maintaining the total energy of the system. In all our simulations, we focus on a temperature of \SI{300}{\kelvin}. The length of the nanowires in the GK simulations amounted to \SI{42.2}{\angstrom} and to obtain the volume, we compute the non-overlapping volume around all of its atoms based on element-specific van-der-Waals spheres of the energy-optimized structures.

\subsection{Cepstral analysis}
Cepstral analysis serves as an efficient technique to evaluate Green-Kubo integrals.\cite{ercole_accurate_2017,ercole_sportran_2022} It is based on the power spectrum $S(\nu)$ of the heat flux, which is given by

\begin{equation}
    \label{e_power_spect}
    S(\nu) = \int_{-\infty}^\infty dt e^{-i2\pi \nu t} \left< J(t) J(0) \right>.
\end{equation}

\noindent Apart from prefactors, the power spectrum at a frequency $\nu=0$ has the same form as equation~\eqref{e_therm_cond} for the thermal conductivity. Therefore, the focus is on analyzing the low-frequency part of the spectrum. An appropriate cutoff frequency has to be chosen, the spectrum is resampled and an inverse discrete Fourier transform of its logarithm is formed to create a so-called \enquote{cepstrum} \cite{cepstrum}. 

This cepstrum is employed to obtain the value of $S(\nu=0)$, a step where we effectively apply a low-pass filter by only using the first few cepstral coefficients $C_n$. The optimal number of cepstral coefficients $P^*$ is chosen objectively by minimizing the second-order Akaike information criterion, AIC$_c$ \cite{burnham_multimodel_2004}:
\begin{equation}
    \label{e_aicc}
    \mathrm{AIC}_c = -2 \log(\mathcal{L}(\theta)) + 2P+\frac{2P(P+1)}{N-P-1}.
\end{equation}

\noindent Here, $P$ is the number of cepstral coefficients, $N$ is the number of candidate models and $\mathcal{L}(\theta)$ is the maximum likelihood function for the model parameters $\theta$. The result for each number of cepstral coefficients represents its separate \enquote{model}. Ercole et al. \cite{ercole_accurate_2017} derived the expression for the maximum log likelihood of the cepstrum as

\begin{equation}
    \label{e_log_likelihood}
    2\log(\mathcal{L}(C_n)) = -\frac{N}{\sigma_l^2} \sum_{n=P}^{N/2}C_n^2,
\end{equation}

\noindent where $\sigma_l^2$ is the variance of all $N$ models. 

The thermal conductivity is then obtained from the optimal number of cepstral coefficients $P^*$ as

\begin{equation}
\label{e_kappa_cepstral_original}
    \kappa_{\mathrm{c}}(P^*) =
     \frac{V}{2k_\mathrm{B} T^2}
    \exp \left[ C_0 + 2 \sum_{n=0}^{P^*-1}C_n-L_0 \right] ,
\end{equation}
where $L_0=\psi(N_f)-\log(N_f)$ is obtained from the number of fluxes $N_f$ used to sample the power spectrum, $\psi$ being the digamma function.
A variance of the thermal conductivity can be estimated with\cite{ercole_sportran_2022}

\begin{equation}
    \label{e_variance_cepstral_original}
    \sigma^2(\kappa_\mathrm{c}(P^*)) = \sigma_0^2 \kappa_c^2 \frac{4P^*-2}{N},
\end{equation}
\noindent where $\sigma_0^2=\psi'(N_f)$ is obtained from the trigamma function $\psi'$.

To improve the consistency of the approach in case several local minima exist for the AIC$_\textrm{c}$, we additionally suggest employing model averaging, a technique commonly used in model selection \cite{burnham_multimodel_2004}. For this, we attribute weights $w_i$ to each model (each number of cepstral coefficients)
\begin{equation}
    \label{e_akaike_weights}
    w_i = \frac{\exp(-\frac{1}{2}\Delta_i)}{\sum_{j=1}^{N} \exp(-\frac{1}{2}\Delta_j)} ,
\end{equation}

\noindent with $\Delta_i = \mathrm{AIC}_{ci}-\min \mathrm{AIC}_c$. The weight distribution is illustrated in Figure~\ref{f_cepstral_at_work}c and the total thermal conductivity from cepstral analysis with model averaging (MA) $\kappa_{\mathrm{cepstral,MA}}$ is predicted using

\begin{equation}
\label{e_kappa_cepstral}
    \kappa_{\mathrm{MA}} =
    \sum_{i=1}^{N} w_i \kappa_\mathrm{c}(P_i).
\end{equation}

\noindent Here, $\kappa_\mathrm{c}(P_i)$ represents the thermal conductivity for a single number of cepstral coefficients as evaluated with equation~\eqref{e_kappa_cepstral_original}.

To estimate the uncertainty of the approach, we evaluate the variance as \cite{lukacs_model_2010}

\begin{equation}
\label{e_error_cepstral}
    \sigma_{\kappa,\mathrm{MA}}^2(C_n)=
    \sum_{j=1}^N w_j\left[ \sigma^2(\kappa_\mathrm{c}(P_j)))+(\kappa_\mathrm{c}(P_j)-\kappa_\mathrm{MA})^2\right].
\end{equation}

\noindent Here, $\sigma^2(\kappa_\mathrm{c}(P_j)))$ is the variance from one model from equation~\eqref{e_variance_cepstral_original}. Note that the uncertainty estimate is always higher when using model averaging than without it.

\subsection{Uncertainty-based analysis based on KUTE}
KUTE is an approach recently proposed by Otero-Lema et al. \cite{otero-lema_kute_2025} to evaluate GK integrals directly based on uncertainty propagation. To describe the approach, we start by introducing a discrete formulation of the elements of the HFACF ($R_k$) from a single MD run, which is required to perform the integration in equation~\eqref{e_therm_cond} numerically

\begin{equation}
    \label{e_hfacf_discrete}
    R_k^{(A)} = \frac{1}{N-k}\sum_{i=0}^{N-k-1} J_{i+k}^{(A)} J_i^{(A)} ,
\end{equation}

\noindent where $N$ is the length of a segment of the heat flux trajectory with index $A$ and $k$ is the index of the correlation time. The discrete elements of the heat flux $J$ at the actual simulation time indices $i$ and $i+k$ are simplified to only include values for the relevant transport direction for the quasi-1D systems discussed here. However, the formulations can be straightforwardly expanded to the 3D case. In order to perform statistical analysis, the entire heat flux trajectory is folded into $M$ equally long pieces, and the average elements of the HFACF are computed with

\begin{equation}
    \label{e_hfacf_from_pieces}
    \bar{R}_k = \frac{1}{M} \sum_{A=1}^M R_k^{(A)}.
\end{equation}

A trapezoidal scheme is used to numerically obtain the cumulative integral of the HFACF:
\begin{equation}
    \label{e_cumul_kute}
    I_{k} = \frac{\Delta t}{2} \sum_{i=0}^k \pqty*{\bar{R}_i + \bar{R}_{i+1}}.
\end{equation}

\noindent In order to evaluate the uncertainty of this expression, KUTE neglects the correlations between the $\bar{R}_k$ and performs standard propagation \cite{otero-lema_kute_2025, JCGMGUM}:

\begin{equation}
    u_t(I_k) = \frac{\Delta t}{2} \sqrt{\sum_{i=0}^k \pqty*{u^2(\bar{R}_i)+u^2(\bar{R}_{i+1})}},
    \label{e_cumul_unc_kute}
\end{equation}

\noindent where $u(\bar{R}_i)$ is obtained for $N$ simulation steps of each equally long trajectory piece using

\begin{equation}
    u(\bar{R}_i) = \frac{1}{\sqrt{(M(N-k))(M(N-k)-1)}}
    \sqrt{\sum_{A=1}^M \sum^{N-k-1}_{i=0} \left(  \bar{R}_k - J_i^{(A)}J_{i+k}^{(A)}  \right)^2},
    \label{e_hfacf_unc_kute}
\end{equation}

\noindent which is equivalent to equation~(4) in the original KUTE paper \cite{otero-lema_kute_2025}.

This propagated uncertainty is then used to create a weighted average of the integral towards the end of the correlation using

\begin{equation}
    \label{e_avg_kappa_to_end}
    \kappa_i^\mathrm{(WA)} = \frac{\sum_{k=i}^N I_k u^{-2}(I_k)}{\sum_{k=i}^N u^{-2}(I_k)},
\end{equation}

\noindent for which the uncertainty follows as

\begin{equation}
    \label{e_unc_kappa_to_end}
    u(\kappa_i^\mathrm{(WA)}) = \sqrt{\frac{1}{N-i} \frac{\sum_{k=i}^N (\kappa_i-I_k)^2 u^{-2}(I_k)}{\sum_{k=i}^Nu^{-2}(I_k)}}.
\end{equation}

This leads to the conductivity and uncertainty for one GK run. To obtain the final value in the KUTE approach, several independent simulations are carried out and the thermal conductivity for the individual independent simulations is then evaluated again using a weighted average with the uncertainty from equation~\eqref{e_unc_kappa_to_end} using

\begin{equation}
    \label{e_wgt_avg_kappa_kute_final}
    \kappa_i^\mathrm{(KUTE)} = \frac{\sum_\alpha \kappa_{i,\alpha}^\mathrm{(WA)} u^{-2}(\kappa_{i,\alpha}^\mathrm{(WA)})}{\sum_\alpha u^{-2}(\kappa_{i,\alpha}^\mathrm{(WA)})},
\end{equation}

where $\alpha$ is the index over the independent simulations. The corresponding uncertainty is

\begin{equation}
    \label{e_unc_wgt_avg_kappa_kute_final}
    u(\kappa_i^\mathrm{(KUTE)}) = \sqrt{\frac{1}{2} 
    \frac{\sum_\alpha \left[ \kappa_{i}-\left(\kappa_{i,\alpha}^\mathrm{(WA)}\right)^2\right] u^{-2}(\kappa_{i,\alpha}^\mathrm{(WA)})}{\sum_\alpha u^{-2}(\kappa_{i,\alpha}^\mathrm{(WA)})}}.
\end{equation}

We will also discuss an alternative approach, using variance based solely on the individual $M$ pieces,

\begin{equation}
    \label{e_hfacf_unc_simple}
    u_{\mathrm{ACF}}(\bar{R}_i) = \sqrt{\frac{1}{M(M-1)} \sum_{A=1}^M \left( R_i^{(A)} - \bar{R}_i \right)^2}.
\end{equation}

One key approximation of the original proposed KUTE approach is that covariances are neglected when the uncertainty of the conductivity integral in equation~\eqref{e_cumul_unc_kute} is propagated. However, neighboring values of $J_{i+k}^{(A)}J_{i}^{(A)}$ will be highly correlated, which will also introduce correlations between the $\bar{R}_k$. Therefore, we propose an extension by introducing the full covariance matrix for the elements of the HFACF based on the average over $M$,

\begin{equation}
    \mathrm{cov}(\bar{R}_i,\bar{R}_j) = 
    \frac{1}{M(M-1)} \sum_{A=1}^M \left[ \left( R_i^{(A)} - \bar{R}_i \right)\right. \left.\left( R_j^{(A)} - \bar{R}_j\right) \right],
\end{equation}

\noindent and we use the simpler Euler integration method to replace equation~\eqref{e_cumul_kute}:

\begin{equation}
    \label{e_cumul_euler}
    I_k = \Delta t \sum_{i=0}^k \bar{R}_i.
\end{equation}

\noindent The effect of this simplification can be expected to be minimal, since it only alters the weights of the terms with $i=0$ and $i=k$. The uncertainty of equation~\eqref{e_cumul_euler} is then computed, without neglecting the correlations, as

\begin{equation}
    \label{e_cumul_unc_cov}
    u_E^\mathrm{(cov)}(I_k) = \Delta t  \sqrt{  \sum_{i=0}^ku_{\mathrm{ACF}}^2(\bar{R}_i)+2\sum_{i=1}^{k} \sum_{j=i+1}^{k} \mathrm{cov}(\bar{R}_i,\bar{R}_j) }.
\end{equation}

\section{Results and discussion}

\subsection{Low-conductivity ZB-phase nanowire}
\begin{figure}
    \centering
    \includegraphics[width=1.0\linewidth]{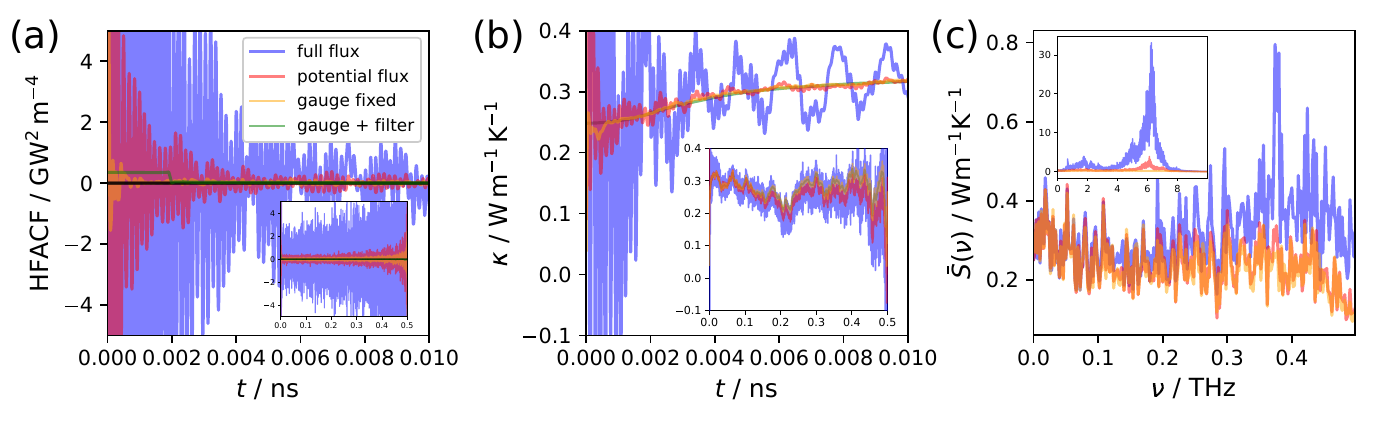}
    \caption{Comparison of different noise-reduction techniques for the ZB nanowire: using the full flux as in equation~\eqref{e_hflux} (blue line), removing the convective term leaving only the potential flux (red line), exploiting the gauge invariance of the flux as in equation~\eqref{e_gauge} (orange line), and employing a smoothening on the gauge fixed flux based on the lowest frequency in the vibrational density of states (green). Comparisons shown are of (a) the heat flux autocorrelation functions, (b) the thermal conductivity $\kappa$, (c) the power spectrum, where the relevant low time or frequency data are shown dominantly and the full ranges are shown as insets.}
    \label{f_spectral_discussion}
\end{figure}

As mentioned in the introduction, a multitude of approaches have been developed to efficiently process the noisy correlation functions for the evaluation of the thermal conductivity at acceptable time scales. To demonstrate the conventionally applied methods, we show the heat flux autocorrelation function and its integral for the ZB nanowire system in Figure~\ref{f_spectral_discussion}a and \ref{f_spectral_discussion}b respectively. Here, the correlation function was averaged over $11$ independent runs with a simulation time of \SI{1}{\nano\second} each and with a correlation time of \SI{0.5}{\nano\second}. The heat flux was only evaluated every $5$ time steps (with \SI{1}{\femto\second} per time step) to save computation time since our tests revealed that a less frequent sampling does not affect the result for the systems considered in this work. It is clear from Figure~\ref{f_spectral_discussion} that the thermal conductivity shows large fluctuations based on correlation length, even after averaging over a substantial amount of simulation time. 

Without averaging, the fluctuations become even more pronounced, making it difficult to estimate the thermal conductivity (see figures~\ref{SI-f_spectral_discussion_0} through \ref{SI-f_spectral_discussion_10} for the $11$ independent runs). In much of the literature, this has been addressed by manually selecting a primary plateau region. This selection is based on the fact that after a certain amount of time, the heat flux is completely uncorrelated. Conventionally, the specific correlation time where this occurs is estimated from the first time the HFACF decays to zero. Thus, this so-called \enquote{first-dip} criterion closes the convergence region right after this first dip is observed. However, this initial region is also prone to noise, making consistent evaluation  difficult. Especially for low conductivity materials, this dip to zero can be very difficult to identify, as shown in Figure~\ref{f_spectral_discussion}a. Hence, specific methods are needed to reduce the level of noise and enable a more objective analysis.

We begin by applying the noise-reduction techniques to the HFACF itself. Since we do not expect ionic transport, it is reasonable to neglect the convective component \cite{Nagoya2022, Zhang2023}. This exclusion leads to a substantial reduction in noise without drastically affecting the resulting integral. Gauge fixing, as defined in equation~\eqref{e_gauge}, has an even more pronounced effect on reducing the high frequency noise, seen especially well in the conductivity integral at low correlation times in  Figure~\ref{f_spectral_discussion}. As another noise-reduction technique, we additionally employed a moving average on either the mirrored HFACF or its integral. The averaging window was chosen based on the inverse lowest frequency peak in the vibrational density of states (VDOS) as suggested by Knoop et al.\cite{knoop_ab_2022}. We computed the VDOS from the velocity autocorrelation function of one of the molecular-dynamics simulations which is shown in Figure~\ref{SI-f_vdos_ZB}, where the lowest peak was identified at \SI{0.2578}{\tera\hertz}. In Figure~\ref{f_spectral_discussion}, we clearly see that for the smoothened HFACF the values close to $t=0$ are too dominant to extract meaningful information beyond a mere observation of the window width of \SI{38.8}{\pico\second}. However, for the thermal conductivity integral, almost all visible high frequency noise was successfully removed by this smoothening procedure. Despite this, none of these techniques reduce the low-frequency oscillations of the thermal conductivity integral substantially.

This becomes even clearer when analyzing the power spectrum of the heat flux shown in Figure~\ref{f_spectral_discussion}c. There, it is evident that the spectrum in the low-frequency region essentially stays the same regardless of whether or not the convective part of the flux is discarded or gauge fixing is utilized. These noise-reduction techniques drastically reduce the magnitude of the high-frequency peaks but do not lead to beneficial effects regarding the low-frequency noise, which is the primary issue when converging the thermal conductivity. Due to its simplicity, we will always discard the convective flux in the results going forward.

\begin{figure}
    \centering
    \includegraphics[width=1.0\linewidth]{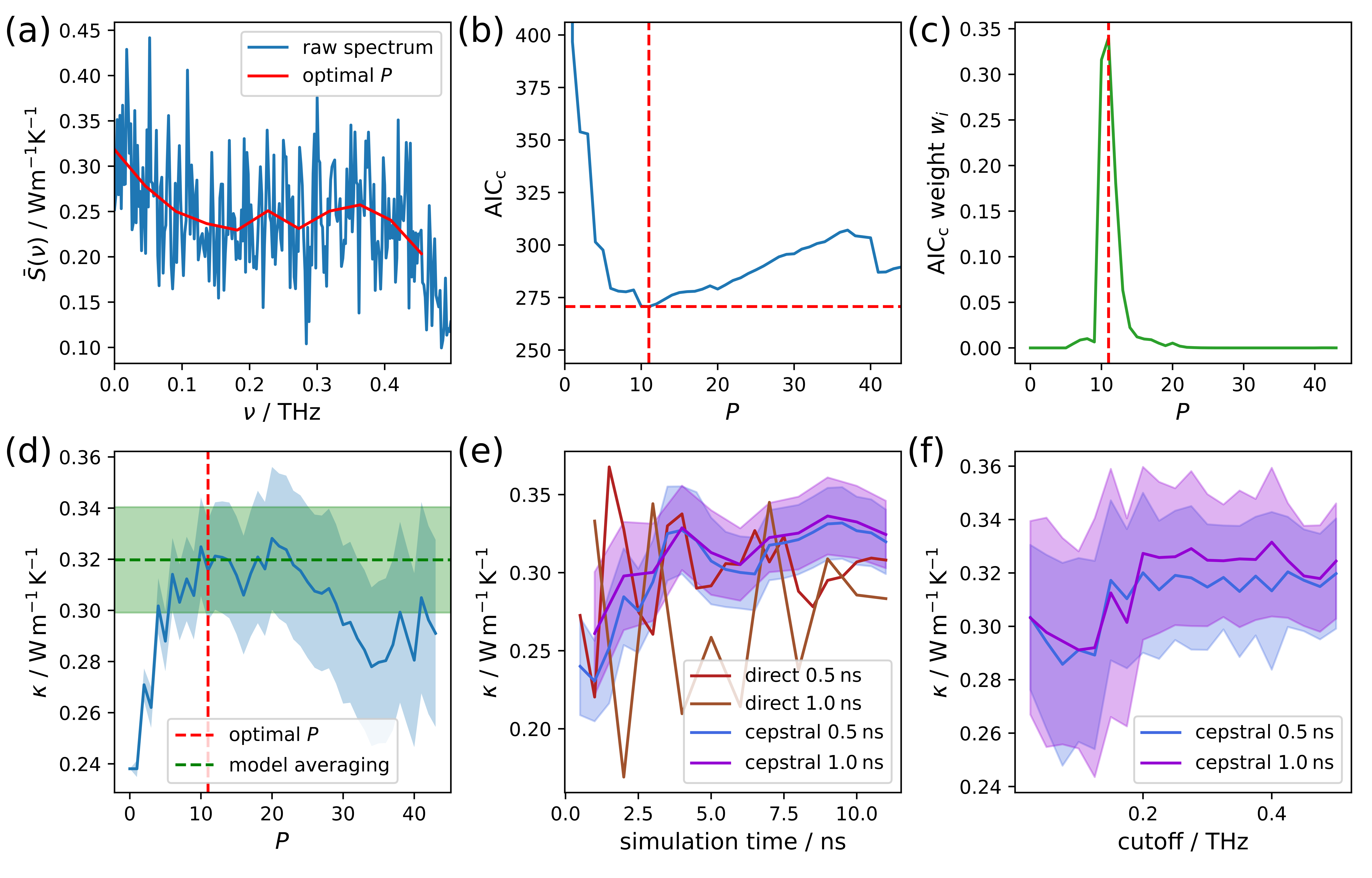}
    \caption{Illustration of the GK evaluation procedure for the ZB nanowire. (a) Low-frequency power spectrum of the HFACF scaled to thermal conductivity units. The red curve indicates the effective smoothing of the spectrum when evaluating the thermal conductivity using the objectively chosen optimal number of cepstral coefficients $P$. (b) Second-order Akaike information criterion (AIC$_c$) as a function of the number of cepstral coefficients. The red dashed lines indicate the minimum. (c) Akaike weight distribution for the range of cepstral coefficients to be used for model averaging. (d) Thermal conductivity $\kappa$ as a function of the number of cepstral coefficients. The resulting values using the optimal number of cepstral coefficients and model averaging are indicated with the red and green dashed lines. (e) Time convergence curves for the thermal conductivity using the direct method and the cepstral analysis. The curves are given for different correlation lengths. (f) Cutoff dependence of the cepstral analysis using the full simulation time.}
    \label{f_cepstral_at_work}
\end{figure}

The primary function of cepstral analysis is to treat precisely this low-frequency region where the denoising techniques fail. To illustrate the effectiveness of this approach, we refer to Figure~\ref{f_cepstral_at_work}, where we illustrate the thermal conductivity evaluation procedure for the ZB nanowire. The low-frequency power spectrum in Figure~\ref{f_cepstral_at_work}a features the actual conductivity value from $S(\nu=0)$, which is the same as the solution for the direct integral before denoising the spectrum.  As only the low-frequency region is of interest, Ercole et al. \cite{ercole_accurate_2017,ercole_sportran_2022} suggest setting a frequency cutoff after the first prominent feature in the spectrum. However, since we do not include the convective contribution causing the higher frequency features, we set the cutoff to a relatively low value of \SI{0.5}{\tera\hertz}. 

To denoise the power spectrum, we first form the $\mathrm{AIC}_c$ as a function of the number of cepstral coefficients and obtain its minimum, as shown in Figure~\ref{f_cepstral_at_work}b. The filtered spectrum based on that optimal number of cepstral coefficients is then shown in Figure~\ref{f_cepstral_at_work}a, with a clearly much reduced level of noise.

It becomes clear that the $\mathrm{AIC}_c$ has a meaningful flat area and in many simulations it features several local minima. The global minimum can easily change with increased simulation time, which is why we do not base our final thermal conductivity result on the minimum alone but rather employ model averaging over several results for different numbers of cepstral coefficients as defined in equation~\eqref{e_kappa_cepstral}.

The results from this model averaging and from several numbers of cepstral coefficients are shown in Figure~\ref{f_cepstral_at_work}d. There, the optimal number of cepstral coefficients is located at the edge of a plateau after which increasing the number of cepstral coefficients does not change the thermal conductivity too severely. Based on our observations, this is precisely the situation where cepstral analysis performs well. This is also reflected in the simulation time convergence curves in Figure~\ref{f_cepstral_at_work}e, where large fluctuations can be observed for the direct integral while the results from cepstral analysis are much more stable and converge reasonably quickly. Additionally, for the cepstral analysis the thermal conductivity values are more similar for different correlation lengths. Note that the direct evaluation here was carried out using the \enquote{first dip} criterion, making the value dependent on a subjective decision and thus potentially influenced by the prejudices of the person making the choice. If the thermal conductivity is not obtained using this criterion, and is instead evaluated after a fixed point in the correlation length, the convergence behavior is far worse, as can be seen in Figure~\ref{SI-f_euler_convergence_ZB}. We also had to choose a value of the frequency cutoff during this analysis. However, in Figure~\ref{f_cepstral_at_work}f, we can see that this choice is of limited significance considering the estimated uncertainty.

\subsection{High-conductivity WZ-phase nanowire}
As promising as the cepstral method looks based on the previous example, for some systems, in particular for those featuring a higher thermal conductivity, its performance is less satisfactory. This is not without precedent, as already in the original work by Ercole et al. \cite{ercole_accurate_2017} the performance for MgO was significantly worse than for the other example systems. There, the authors argued that more reasonable values can be obtained using twice the number of cepstral coefficients. Our aim here is to discuss the origin of the problem and possible solutions to it using a relatively extreme example. It must also be pointed out that, to achieve fully quantitative values for higher-conductivity systems, careful convergence tests with respect to the size of the simulation box can be required. Here we focus on optimizing convergence for a given simulation box, to be consistent with respect to the number of atoms and size of the system with the ZB-phase example, and to avoid a combinatorial explosion of situations with different convergence issues. For instance, a longer simulation box leads to longer tails in the correlation function and to the need to reassess convergence. If an extrapolation to infinite size is desired, an approach based on the lifetimes of the heat carriers \cite{knoop_ab_2022} exists that can reduce computation time in practice.

\begin{figure}
    \centering
    \includegraphics[width=0.5\linewidth]{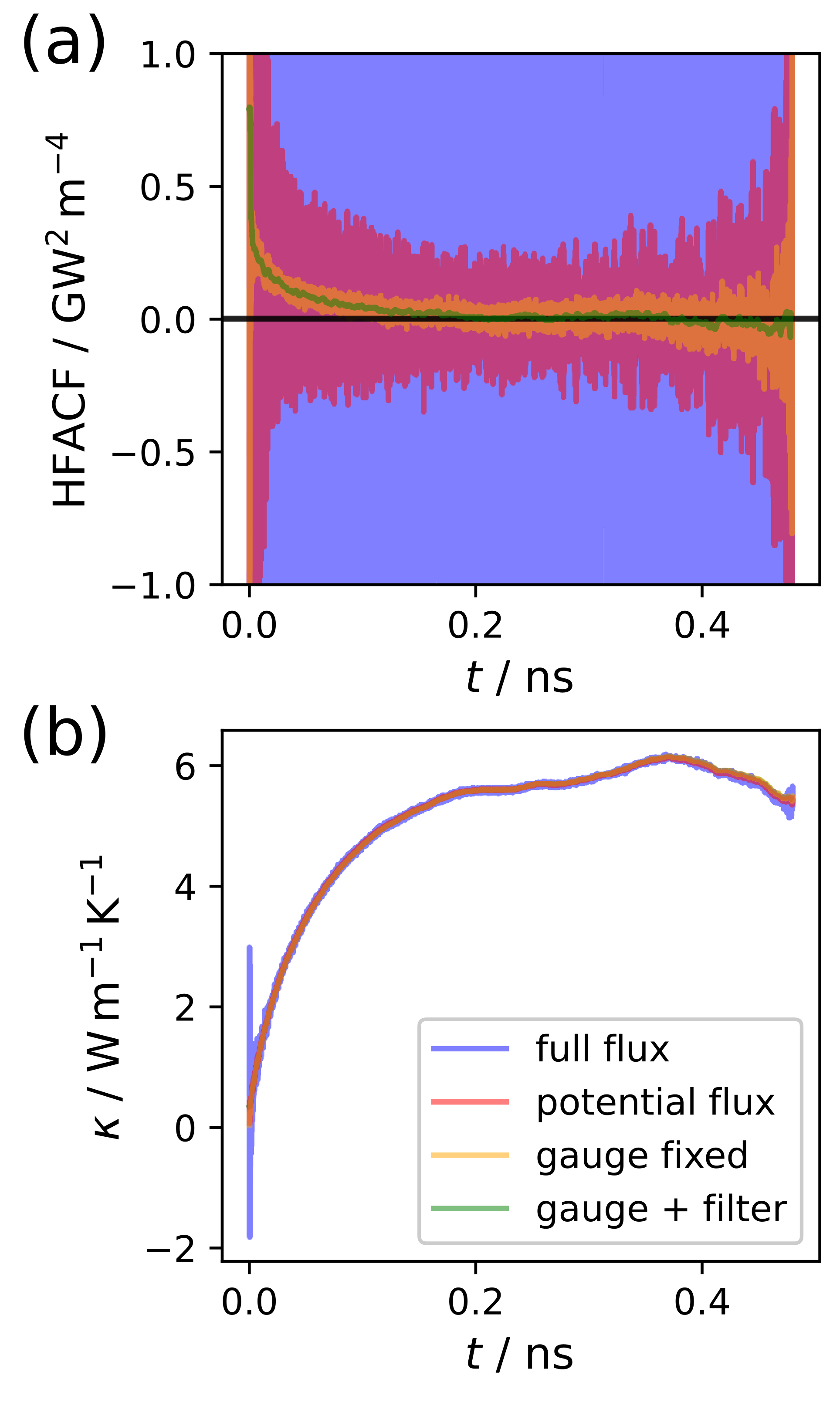}
    \caption{Comparison of different noise-reduction techniques for the WZ nanowire: using the full flux as in equation~\eqref{e_hflux}, removing the convective term leaving only the potential flux, exploiting the gauge invariance of the flux as in equation~\eqref{e_gauge}, and for additionally applying a smoothening filter with a window width corresponding to the lowest peak in the vibrational density of states. Comparisons shown are for (a) the heat flux autocorrelation functions, and (b) the thermal conductivity $\kappa$. }
    \label{f_spectral_discussion_WZ}
\end{figure}

To illustrate the central issue, we show the analysis for a higher-thermal-conductivity WZ-phase nanowire starting with the initial noise-reduction techniques in Figure~\ref{f_spectral_discussion_WZ}. In this case, the total simulation time (a total of \SI{75}{\nano\second} accumulated over 55 independent runs) is much higher. Gauge fixing also shows its effectiveness for reducing the high frequency noise. This reveals the much slower decay of the correlation function in Figure~\ref{f_spectral_discussion_WZ}a which reaches zero only after approximately \SI{200}{\pico\second}. The slow decay of the HFACF had been shown previously for \ch{Si} nanowires \cite{Zhou2017}, where lower diameters led to even larger required correlation times. The long decay time of the correlation is also an indicator of the presence of high-lifetime phonons in the system, making long simulations necessary to achieve convergence. Similarly to the procedure for the ZB nanowire, we also applied a smoothening filter in the form of a running average with a window width corresponding to the lowest frequency peak in the VDOS (\SI{0.333}{\tera\hertz}, see Figure~\ref{SI-f_vdos_WZ}). For this higher-conductivity nanowire, applying a filter with this objective selection criterion leads to an excellent smoothening behavior also for the HFACF. However, considering the very similar position of the lowest-frequency peak in both of the considered systems and their completely different convergence behaviors, it is likely that an universally applicable criterion to choose the window width would also need to include an approximation of the decay time of the HFACF. Despite the effectiveness of gauge fixing and further smoothening for the HFACF, the direct integral in Figure~\ref{f_spectral_discussion_WZ}b barely shows any differences based on whether or not the noise-reduction techniques were applied.

Despite the high-frequency noise clearly not being the primary issue here, we do want to use this opportunity to briefly discuss a relatively uncertain aspect regarding the ideal strategy to increase simulation time: is it better to add more independent simulations or is it better to utilize fewer but long simulations? In the supplementary information [Figures~\ref{SI-f_noise_choice} and \ref{SI-f_noise_choice_kappa}] we show the HFACF and its integral when we increase the number of independent simulations vs. when we just increase their length. It can be seen clearly that, for this particular example, the noise decreases very similarly regardless of the  strategy employed. Based on our experience the deciding factor is merely the total simulation time. Despite the seeming irrelevance of this choice for our systems, for less well-behaved systems with several unique local energy minima a larger number of independent simulations is still expected to be more beneficial.

\begin{figure}
    \centering
    \includegraphics[width=1.0\linewidth]{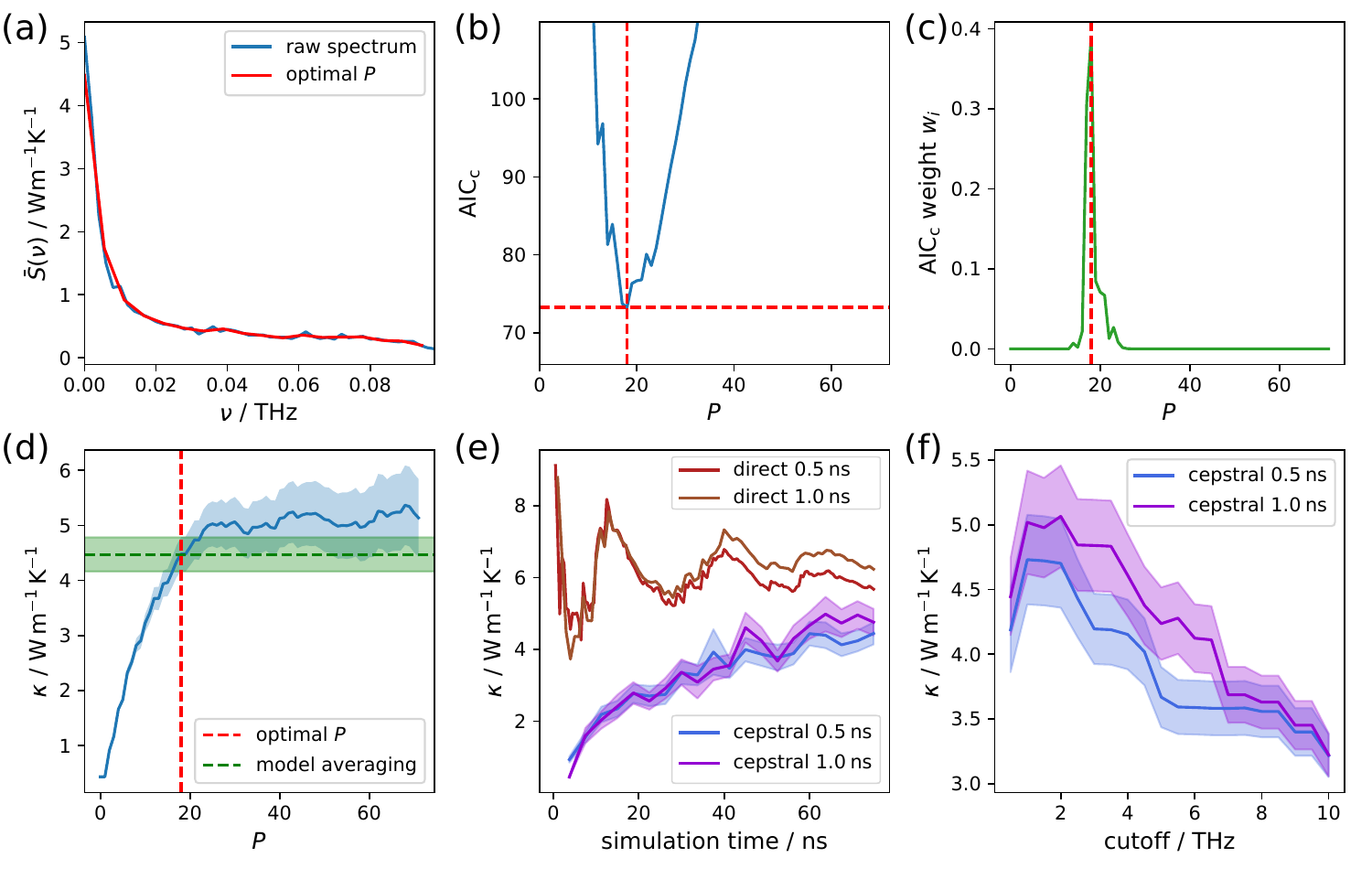}
    \caption{Issues with the cepstral analysis for the higher thermal conductivity WZ nanowire. (a) Low-frequency power spectrum of the HFACF scaled to thermal conductivity units. The red curve indicates the effective smoothing of the spectrum when evaluating the thermal conductivity using the objectively chosen optimal number of cepstral coefficients $P$. (b) Second-order Akaike information criterion (AIC$_c$) as a function of the number of cepstral coefficients. The red dashed lines indicate the minimum. (c) Akaike weight distribution for the range of cepstral coefficients to be used for model averaging. (d) Thermal conductivity $\kappa$ as a function of the number of cepstral coefficients. The resulting values using the optimal number of cepstral coefficients and model averaging are indicated with the red and green dashed lines. (e) Time convergence curves for the thermal conductivity using the direct method and the cepstral analysis. The curves are shown for different correlation lengths. (f) Cutoff dependence of the cepstral analysis using the full simulation time.}
    \label{f_cepstral_not_working}
\end{figure}

In Figure~\ref{f_cepstral_not_working}a, the issue with cepstral analysis is apparent in the low-frequency power spectrum. There is barely any noise and the thermal conductivity result for the optimal number of cepstral coefficients severely underestimates the thermal conductivity as evaluated from $S(\nu=0)$. The choice of the number of cepstral coefficients heavily influences the result, as can be seen in Figure~\ref{f_cepstral_not_working}d. It is clear that cepstral analysis leads to a similar value as the direct approach when choosing a larger number of cepstral coefficients. Similar observations were made in the work of Ercole et al. \cite{ercole_accurate_2017}, where a number of cepstral coefficients twice as high as the general recipe would prescribe also led to a good result for MgO. However, in Figures~ \ref{f_cepstral_not_working}b and \ref{f_cepstral_not_working}c we see a relatively clear minimum of the Akaike information criterion with no substantial local minimum at a higher number of coefficients. The convergence curves in Figure~\ref{f_cepstral_not_working}e indicate that cepstral analysis would eventually reach similar values as the direct approach, but after a prohibitively long simulation time. In Figure~\ref{f_cepstral_not_working}f we see that the choice of the cutoff is extremely relevant in this instance. However, since there is no plateau, it is difficult to create an objective criterion for its selection.

Based on these observations, it appears clear that in this case the cepstral analysis based on the Akaike information criterion is not suitable for obtaining the thermal conductivity for these higher-thermal-conductivity materials, where the direct approach leads to a faster convergence behavior than cepstral analysis. However, the required simulation time becomes prohibitively long. Therefore, in the following we also explore alternative approaches using an uncertainty-based analysis.

\subsection{Uncertainty-based approach for higher-conductivity systems}
One of the key benefits of the cepstral analysis was that it inherently provides an error estimate of the resulting conductivity value. A different approach, KUTE \cite{otero-lema_kute_2025}, has recently been developed that focuses on the uncertainty in the process of direct integration. However, this has not been directly used for heat transport simulations yet and only discussed in the context of electric conductivity, viscosity and diffusion coefficients. The general concept of the approach was introduced in the method section.

\begin{figure}
    \centering
    \includegraphics[width=0.5\linewidth]{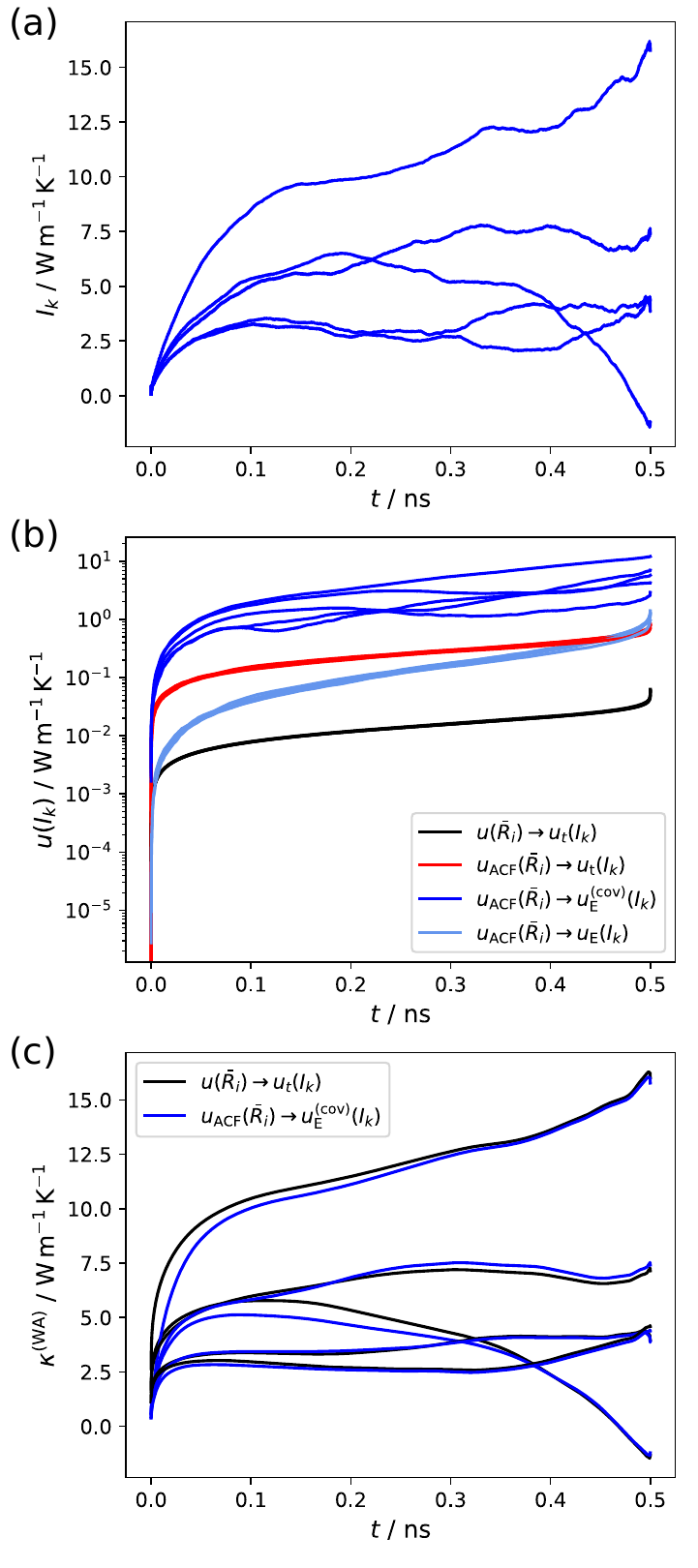}
    \caption{Comparison of analysis methods based on five 5-ns GK simulations of the WZ nanowire. (a) Direct integration of a $M = 10$-times-folded HFACF for the 5 independent runs. (b) Comparison of different uncertainty metrics propagated from the HFACF uncertainties defined in equations~\eqref{e_hfacf_unc_kute} ($u(\bar{R}_i)$) and \eqref{e_hfacf_unc_simple} ($u_\mathrm{ACF}(\bar{R}_i)$) into the integral uncertainties from equations~\eqref{e_cumul_unc_kute} ($u_t(I_k)$) and \eqref{e_cumul_unc_cov} excluding ($u_E(I_k)$) or including ($u^\mathrm{(cov)}_E(I_k)$) the covariance contributions. (c) Uncertainty-weighted averages as defined in equation~\eqref{e_avg_kappa_to_end} using the uncertainty $u_t(I_k)$ (black) as suggested by the KUTE approach or the uncertainty $u^\mathrm{(cov)}_E(I_k)$ (blue) including the full covariance matrix.}
    \label{f_uncertainty_discussion}
\end{figure}
We begin by applying the KUTE approach to the WZ nanowire using 5-ns GK simulations with $M=10$, see equation~\eqref{e_hfacf_from_pieces}, resulting in correlation lengths of \SI{0.5}{\nano\second}. Performing five independent of such runs, Figure~\ref{f_uncertainty_discussion}a illustrates how substantial discrepancies are observed even after a significant simulation time. This underscores the necessity of an uncertainty-based framework to assign appropriate weights to the data.

In Figure~\ref{f_uncertainty_discussion}b, the corresponding uncertainty from equation~\eqref{e_cumul_unc_kute} is shown as a black line. While the uncertainty qualitatively behaves reasonably well in that it leads to high values at the tail end where data is scarce, it also leads to extremely low quantitative uncertainty values considering the enormous spread of the independent simulations, Figure~\ref{f_uncertainty_discussion}a. If we then perform the weighted average as in equation~\eqref{e_avg_kappa_to_end} we obtain integral curves, as shown by the blue lines in Figure~\ref{f_uncertainty_discussion}c, with the initial rise shifted to lower correlation times leading to a slightly larger plateau. However, the propagated uncertainties from equation~\eqref{e_unc_kappa_to_end} are still very small in the range of $10^{-2}$ to $10^{-3}$~\SI{}{\watt\per\meter\per\kelvin} as can be seen in Figure~\ref{SI-f_uncertainty_kute_individual}.

One of the reasons for this low uncertainty originates from equation~\eqref{e_hfacf_unc_kute}, where it was evaluated for the HFACF including the length $N-k$ from the calculation of the correlation function. Therefore, we also employ equation~\eqref{e_hfacf_unc_simple} to obtain the uncertainty solely based on the $M$ pieces, whose effect can be seen in the uncertainties shown as the red area in Figure~\ref{f_uncertainty_discussion}b. They are significantly larger, but still much less than what would be expected of a quantitative error from the spread in Figure~\ref{f_uncertainty_discussion}a.

The uncertainties including the covariance contribution are shown as dark blue lines in Figure~\ref{f_uncertainty_discussion}b. While the curves are qualitatively similar to those from equations~\eqref{e_cumul_unc_kute} and \eqref{e_hfacf_unc_simple}, their magnitudes are consistent with the large spread of the $I_k$ values [Figure~\ref{f_uncertainty_discussion}a] indicating that equation~\eqref{e_cumul_unc_cov} provides a more appropriate quantitative error metric. Interestingly, only in this case do the  uncertainties of the independent runs show significant deviations from each other. To underline the importance of the covariance contribution, we furthermore show the uncertainty obtained by neglecting the second term under the square root in equation~\eqref{e_cumul_unc_cov}  (light blue curves labeled $u_E(I_k)$ in 
Figure~\ref{f_uncertainty_discussion}b). Thereby, uncertainties of a similar order of magnitude as those obtained by the straightforward application of equation~\eqref{e_hfacf_unc_simple} (red line).

All the methods to obtain the uncertainty show the largest values towards the end of the correlation time. This is an important characteristic for further uncertainty-based evaluation, as at that point, the statistics of the computed correlation functions in equation~\eqref{e_hfacf_discrete} become quite poor.

\begin{figure}
    \centering
    \includegraphics[width=0.5\linewidth]{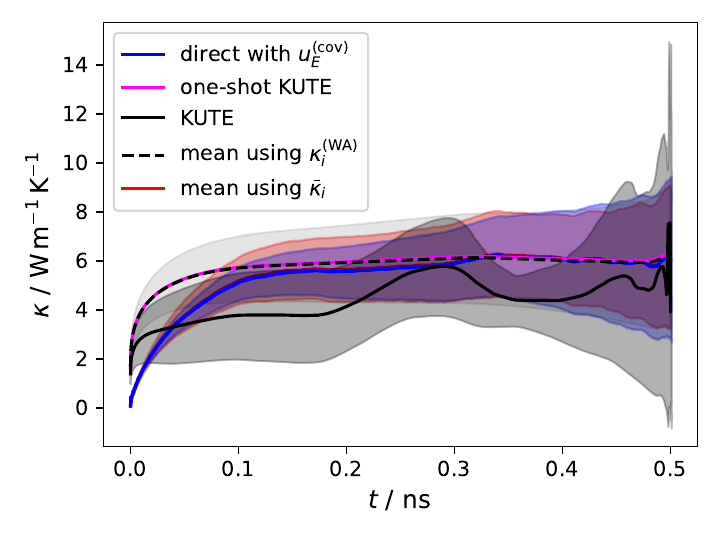}
    \caption{Comparison of different analysis techniques of the thermal conductivity with their uncertainty prediction. This includes the direct analysis of the thermal conductivity including the covariance contribution of the uncertainty $u^\mathrm{(cov)}_E(I_k)$ (blue line), an one-shot KUTE approach where the weighted average $\kappa^\mathrm{(WA)}$ was performed once over all simulation data with the very small invisible uncertainty $u(\kappa^\mathrm{(WA)})$ from equation~\eqref{e_unc_kappa_to_end} (magenta line), the KUTE approach as prescribed from equation~\eqref{e_wgt_avg_kappa_kute_final} (black solid line) with its uncertainty from equation~\eqref{e_unc_wgt_avg_kappa_kute_final} (dark gray shaded area), and using an arithmetic mean over the weighted averages $\kappa^\mathrm{(WA)}$ (black dashed line) or over the untreated conductivity integral $\bar{\kappa}$ (red line) from the data of all independent simulations with the standard deviation as the error metric (light gray or red shaded area respectively)}
    \label{f_euler_kute_comparison}
\end{figure}

In the spirit of the KUTE approach, we can also obtain the weighted average using the uncertainties including covariances, as shown in Figure~\ref{f_uncertainty_discussion}c. Despite the vast difference in the original uncertainties, the averaged curves appear to be similar. In both cases, the uncertainties from equation~\eqref{e_unc_kappa_to_end} are very small and lose the characteristically high value towards the end of the correlation time that was present before performing the weighted mean. 

In the prescribed KUTE approach, the final step is to perform another weighted average over the five independent simulations, equation~\eqref{e_wgt_avg_kappa_kute_final}. The resulting thermal conductivity is shown as the solid black curve in Figure~\ref{f_euler_kute_comparison} with its uncertainty, equation~\eqref{e_unc_wgt_avg_kappa_kute_final}, represented by the dark gray area. The obtained thermal conductivity is not fully stable and even exhibits inconsistent behavior at larger correlation times, which can be traced to the erratic behavior of the uncertainties from the independent weighted average (see Figure~\ref{SI-f_uncertainty_kute_individual}). Nevertheless, the uncertainty obtained from equation~\eqref{e_unc_wgt_avg_kappa_kute_final} results in reasonably quantitative values, as it primarily reflects the spread of the independent values rather than the uncertainties used as weights. 

For comparison, the arithmetic average over all individual simulations evaluated with equation~\eqref{e_avg_kappa_to_end} yields a more stable thermal conductivity (dashed black line with light gray error bars obtained from the standard error in Figure~\ref{f_euler_kute_comparison}). The result matches a one-shot KUTE evaluation, where the five independent simulations are combined directly into $M=50$ pieces and the final conductivity is obtained from equation~\eqref{e_avg_kappa_to_end}, shown as the magenta line in Figure~\ref{f_euler_kute_comparison}, in which case the uncertainty loses its quantitative character. 
A similar trend can also be seen for the blue curve, which represents a direct one-shot evaluation including the full covariance contribution. In this case, due to not using equation~\eqref{e_avg_kappa_to_end}, the curve shows a slower convergence behavior, with a small uncertainty at the beginning which increases toward the end, a behavior that was not present for the KUTE based approaches. For illustrative purposes, we additionally perform a comparison to the simple arithmetic mean over the thermal conductivity integrals for the same correlation length, with an error provided by the standard deviation over these runs. This traditional evaluation approach leads to similar values for the thermal conductivity and its uncertainty but does not allow the consideration of intermediate uncertainties for each run.

\begin{figure}
    \centering
    \includegraphics[width=1.0\linewidth]{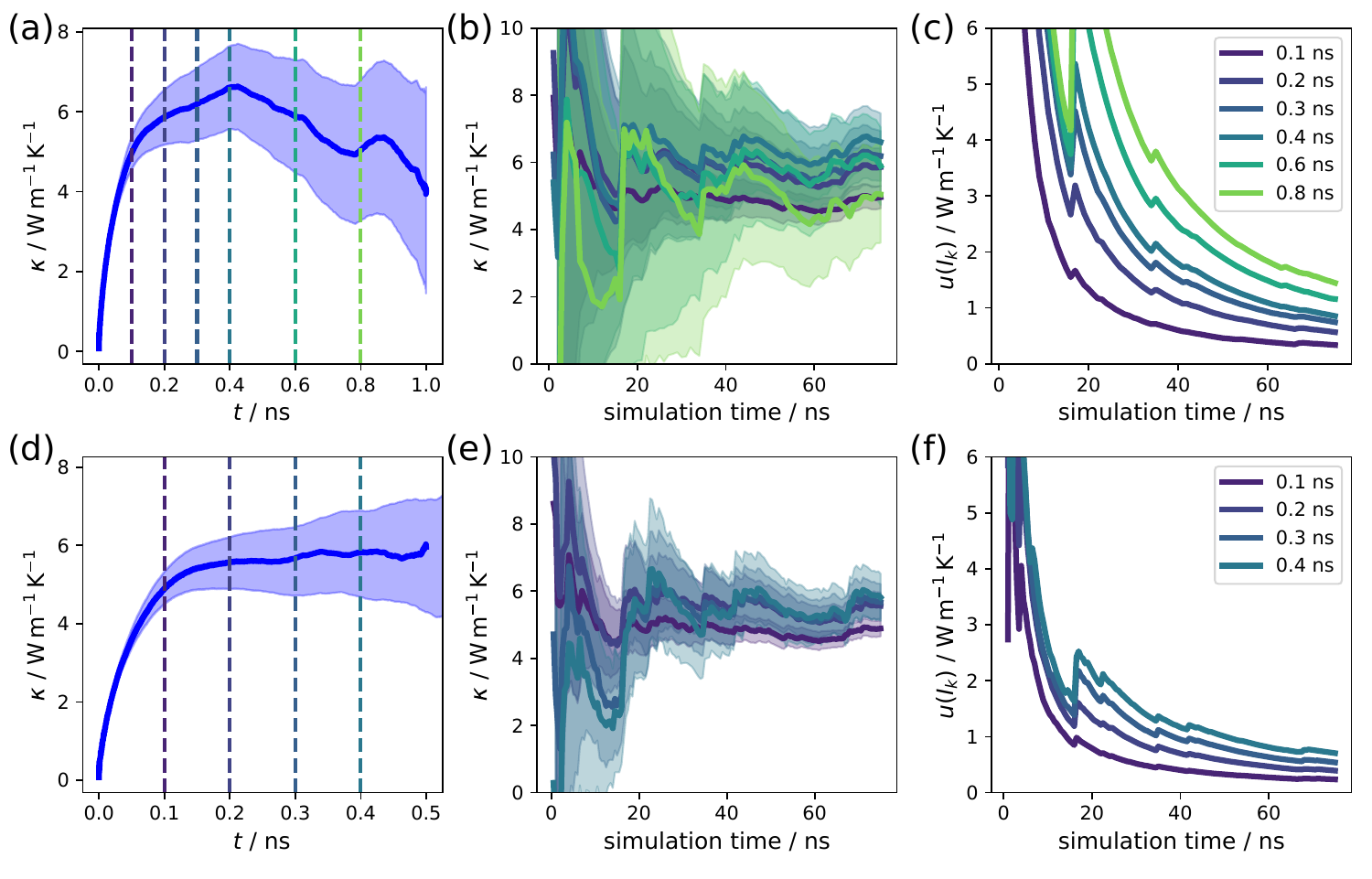}
    \caption{Time convergence of the WZ nanowire with a correlation time of \SI{1}{\nano\second} (a, b, c) or \SI{0.5}{\nano\second} (d, e, f) evaluated over a total simulation time of \SI{75}{\nano\second}. (a, d) Cumulative integrals performed using the Euler method and the corresponding uncertainty for the entire duration. (b, e) Time convergence curves based on values extracted after specific correlation times as indicated in (a,b). The corresponding uncertainties are shown in (c, f). }
    \label{f_euler_convergence}
\end{figure}

Nonetheless, these results indicate that neglecting the covariance matrix can introduce a large error, and that it needs to be considered when relying on error propagation. Even if it is included, it is still necessary to identify the plateau region of the thermal conductivity with respect to correlation time. The availability of an uncertainty metric, however, does provide a better guideline to properly choose said region and a way to estimate the error in the final result. We illustrate this utility in Figure~\ref{f_euler_convergence}, where the evaluation is shown for two different correlation times of 1.0 and \SI{0.5}{\nano\second}. In the latter case, the values were obtained from twice as many individual pieces. It is clear that the longer correlation time features a very large level of noise that was only barely captured using the described evaluation method. Additionally, a correlation time of at least \SI{0.2}{\nano\second} is required to reach a plateau in Figure~\ref{f_euler_convergence}d  that coincides with the correlation time where the HFACF decays to zero in Figure~\ref{f_spectral_discussion_WZ}a. This explains the lower value of the time convergence curves for the extraction at \SI{0.1}{\nano\second} in Figures~\ref{f_euler_convergence}b and e. In Figure~\ref{f_euler_convergence}a and at an extraction point of 0.8 ns, the thermal conductivity value is substantially lower and, in that particular case, the uncertainty becomes extremely large. However, the convergence curves do agree quite well for different correlation-length extraction points when one considers the substantial errors above \SI{1}{\watt\per\meter\per\kelvin}. Figures~\ref{f_euler_convergence}c and f also clearly show that the uncertainty decreases as a function of time. Individual outlier simulations, with anomalously high values, can cause substantial increases in the function leading to a slow decay. If the goal is high accuracy, extremely large simulation times are required.

To also test this approach for the ZB nanowire, we showcase the analogous uncertainty evaluation in Figure~\ref{SI-f_uncertainty_discussion_ZB}. Switching from the trapezoid rule to the simpler Euler integration leads to more substantial differences for the low conductivity ZB nanowire  when compared to the WZ nanowire. However, the propagated uncertainties from both integration methods, even without the contribution of covariances, and just using equation~\eqref{e_hfacf_unc_simple}, are far larger than the differences. This highlights the increasing lack of precision of the thermal conductivity integral with correlation time.

Additionally, for the ZB nanowire, with a very low thermal conductivity, the uncertainty including covariance contributions is extremely large (see Figure~\ref{SI-f_euler_convergence_ZB}) and does not appear to decrease substantially with simulation time. This clearly emphasizes the benefit of the cepstral analysis method for low conductivity materials.

\begin{figure}[htb!]
    \centering
    \includegraphics[width=1.0\linewidth]{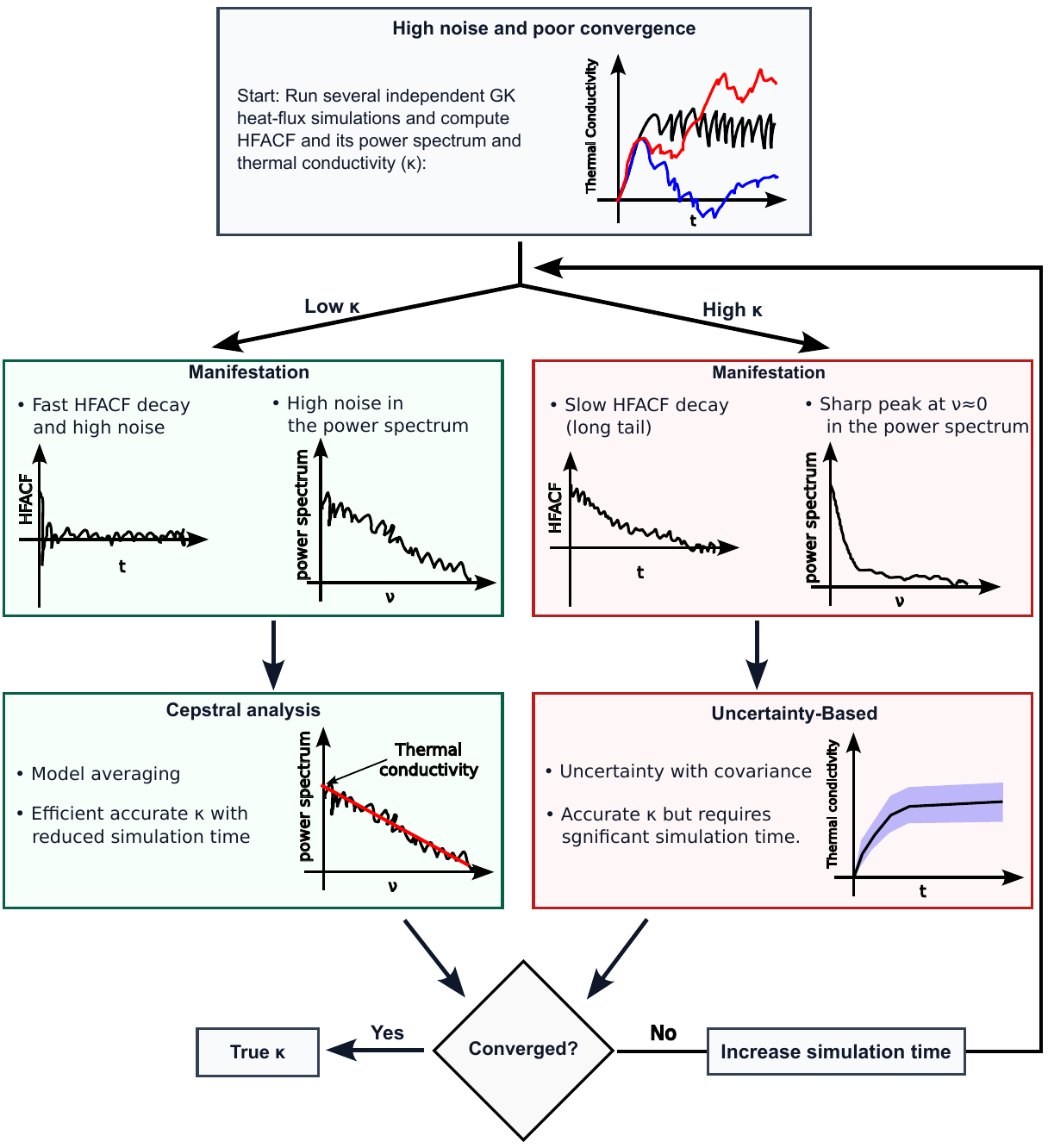}
    \caption{Schematic decision tree illustrating our methodological recommendations.}
    \label{fig:workflow}
\end{figure}

\section{Conclusion}

In this work, we analyzed the application of GK simulations on quasi-1D nanowire systems. We created and validated accurate and, within the context of thin nanowire structures, transferable MACE models for \ch{InAs}. To perform Green-Kubo simulations, we implemented an efficient strategy to properly evaluate the heat flux for MACE models and made the code publicly available \cite{mace-unfolded_2025}.

The GK approach was applied to two nanowire structures from the \ch{InAs} zincblende (ZB) or wurtzite (WZ) phase respectively. We show, that despite their similarities in diameter and in the bulk crystal phase, the ZB nanowire shows a strongly reduced thermal conductivity value. This makes treating Green-Kubo simulations for them with the same analysis techniques difficult while maintaining a low computational profile. 

Cepstral analysis \cite{ercole_accurate_2017} is a promising technique to severely reduce the simulation time required of the difficult-to-converge GK simulations. We expanded on the approach by employing a model-averaging technique leading to an improved convergence behavior. For the low-conductivity ZB nanowire, cepstral analysis is shown to be an efficient solution to reduce the level of noise and therefore the simulation time required. However, we show that the same approach fails for the high-conductivity WZ nanowire, which features a long tail of the heat-flux autocorrelation function (HFACF). The thermal conductivity obtained from cepstral analysis significantly underestimates the correct result due to the difficulty it faces when analyzing sharp peaks of the relatively low-noise power spectrum at zero frequency. In case of observing such a peak, it is necessary to use longer simulation times and perform the analysis with a direct integral to achieve proper convergence. 

To allow for an objective selection of the convergence region, we discuss approaches to evaluating the uncertainty of the conductivity integral. There, we employed the recently proposed KUTE \cite{otero-lema_kute_2025} approach, which is based on uncertainty propagation from the averaged HFACF. However, KUTE features non-quantitative intermediate uncertainties, which in a large part originate from the neglected correlation effects during error propagation. We propose an approach including the covariance contribution within the framework of a simplified Euler integral which features quantitative uncertainties of GK simulations for different correlation times. That uncertainty analysis allows a selection of a value that agrees within the uncertainty as long as the correlation time is long enough to capture the decay of the HFACF. The obtained error is substantial and decays only slowly with increasing simulation time showcasing the limits of the direct evaluation technique.

In summary, while cepstral analysis can severely reduce the required simulation time of GK simulations for low conductivity systems, one should be careful in its application as it can lead to a severe underestimation for materials featuring slow decays of the HFACF. In such a case, it is recommended to employ an uncertainty-based approach considering correlation and a significant number of independent simulations. Figure~\ref{fig:workflow} illustrates these recommendations.

\section{Data availability}
The MACE model, datasets, and Green-Kubo results are publicly available for download at \url{https://doi.org/10.5281/zenodo.16913162}. The MACE heat flux calculator is publicly available at \url{https://github.com/pulgon-project/mace-unfolded}, and the Green-Kubo analysis package can be obtained from \url{https://github.com/pulgon-project/gk-analysis}.

\begin{acknowledgement}
This research was funded in whole or in part by the Austrian Science Fund (FWF) [10.55776/P36129]. For open access purposes, the author has applied a CC BY public copyright license to any author-accepted manuscript version arising from this submission. It was also supported by MCIN with funding from the European Union NextGenerationEU (PRTR-C17.I1) promoted by the Government of Aragon. J.C. acknowledges Grant CEX2023-001286-S funded by MICIU/AEI /10.13039/501100011033.
\end{acknowledgement}

\begin{suppinfo}

The following files are available free of charge.

\begin{itemize}
  \item Supporting Information: Figure~\ref{SI-f_ZB_band_comparison}: Phonon band structures computed with DFT or the MACE model; Figures~\ref{SI-f_vdos_ZB} and \ref{SI-f_vdos_WZ} showing the vibrational density of states for the ZB and WZ nanowires respectively; Figures~\ref{SI-f_spectral_discussion_0} through \ref{SI-f_spectral_discussion_10}: HFACF, its integral and its power spectrum for 11 individual simulations for the zincblende phase; Figures~\ref{SI-f_zero_freq_plot_ZB} and \ref{SI-f_zero_freq_plot_WZ}  showing the zero frequency thermal conductivity extracted from the power spectrum for both phases; Figures~\ref{SI-f_noise_choice} and \ref{SI-f_noise_choice_kappa} illustrate differences in the noise of Green-Kubo simulations based on different strategies to increase simulation time; Figure~\ref{SI-f_uncertainty_kute_individual}: thermal conductivities and uncertainties from the prescribed KUTE approach; Figure~\ref{SI-f_uncertainty_discussion_ZB}: direct integration methods applied to the low conductivity ZB nanowire; Figure~\ref{SI-f_euler_convergence_ZB}: Time convergence from the direct approach using the Euler propagation method
\end{itemize}

\end{suppinfo}

\bibliography{references-2}{}

@ARTICLE{cepstrum,
  author={Childers, D.G. and Skinner, D.P. and Kemerait, R.C.},
  journal={Proc. IEEE}, 
  title={The cepstrum: A guide to processing}, 
  year={1977},
  volume={65},
  number={10},
  pages={1428--1443},
  doi={10.1109/PROC.1977.10747}}

@article{Nagoya2022,
  title = {Autonomous Search for Polymers with High Thermal Conductivity Using a Rapid Green–Kubo Estimation},
  volume = {55},
  ISSN = {1520-5835},
  url = {http://dx.doi.org/10.1021/acs.macromol.1c02267},
  DOI = {10.1021/acs.macromol.1c02267},
  number = {9},
  journal = {Macromolecules},
  publisher = {American Chemical Society (ACS)},
  author = {Nagoya,  Akihiro and Kikkawa,  Nobuaki and Ohba,  Nobuko and Baba,  Takeshi and Kajita,  Seiji and Yanai,  Kazuma and Takeno,  Takanori},
  year = {2022},
  month = apr,
  pages = {3384–3395}
}

@article{Zhang2023,
  title = {Green-Kubo formalism for thermal conductivity with Slater-Koster tight binding},
  volume = {108},
  ISSN = {2469-9969},
  url = {http://dx.doi.org/10.1103/PhysRevB.108.104307},
  DOI = {10.1103/physrevb.108.104307},
  number = {10},
  pages = {104307},
  journal = {Physical Review B},
  publisher = {American Physical Society (APS)},
  author = {Zhang,  Dong-Bo and Li,  Jian-Gao and Ren,  Ying-Hui and Sun,  Tao},
  year = {2023},
  month = sep 
}

@article{togo_distributions_2015,
  title = {Distributions of phonon lifetimes in Brillouin zones},
  volume = {91},
  ISSN = {1550-235X},
  url = {http://dx.doi.org/10.1103/PhysRevB.91.094306},
  DOI = {10.1103/physrevb.91.094306},
  number = {9},
  pages = {094306},
  journal = {Physical Review B},
  publisher = {American Physical Society (APS)},
  author = {Togo,  Atsushi and Chaput,  Laurent and Tanaka,  Isao},
  year = {2015},
  month = mar 
}

@article{Li2014,
  title = {ShengBTE: A solver of the Boltzmann transport equation for phonons},
  volume = {185},
  ISSN = {0010-4655},
  url = {http://dx.doi.org/10.1016/j.cpc.2014.02.015},
  DOI = {10.1016/j.cpc.2014.02.015},
  number = {6},
  pages = {1747-1758},
  journal = {Computer Physics Communications},
  publisher = {Elsevier BV},
  author = {Li,  Wu and Carrete,  Jesús and A. Katcho,  Nebil and Mingo,  Natalio},
  year = {2014},
  month = jun
}

@article{li_thermal_2013,
  title = {Thermal conductivity of bulk and nanowire InAs,  AlN,  and BeO polymorphs from first principles},
  volume = {114},
  ISSN = {1089-7550},
  url = {http://dx.doi.org/10.1063/1.4827419},
  DOI = {10.1063/1.4827419},
  number = {8},
  pages = {183505},
  journal = {Journal of Applied Physics},
  publisher = {AIP Publishing},
  author = {Li,  Wu and Mingo,  Natalio},
  year = {2013},
  month = nov 
}

@article{karim_temperature-dependence_2020,
  title = {Temperature-dependence calculation of lattice thermal conductivity and related parameters for the zinc blende and wurtzite structures of InAs nanowires},
  volume = {43},
  ISSN = {0973-7669},
  url = {http://dx.doi.org/10.1007/s12034-019-2011-1},
  DOI = {10.1007/s12034-019-2011-1},
  number = {1},
  pages = {54},
  journal = {Bulletin of Materials Science},
  publisher = {Springer Science and Business Media LLC},
  author = {Karim,  Hawbash H and Omar,  M S},
  year = {2020},
  month = jan 
}

@article{Schelling2002,
  title = {Comparison of atomic-level simulation methods for computing thermal conductivity},
  volume = {65},
  ISSN = {1095-3795},
  url = {http://dx.doi.org/10.1103/PhysRevB.65.144306},
  DOI = {10.1103/physrevb.65.144306},
  number = {14},
  pages = {144306},
  journal = {Physical Review B},
  publisher = {American Physical Society (APS)},
  author = {Schelling,  Patrick K. and Phillpot,  Simon R. and Keblinski,  Pawel},
  year = {2002},
  month = apr 
}

@article{liang_finite-size_2014,
  title = {Finite-size effects on molecular dynamics interfacial thermal-resistance predictions},
  volume = {90},
  ISSN = {1550-235X},
  url = {http://dx.doi.org/10.1103/PhysRevB.90.075411},
  DOI = {10.1103/physrevb.90.075411},
  number = {7},
  pages = {075411},
  journal = {Physical Review B},
  publisher = {American Physical Society (APS)},
  author = {Liang,  Zhi and Keblinski,  Pawel},
  year = {2014},
  month = aug 
}

@article{Sellan2010,
  title = {Size effects in molecular dynamics thermal conductivity predictions},
  volume = {81},
  ISSN = {1550-235X},
  url = {http://dx.doi.org/10.1103/PhysRevB.81.214305},
  DOI = {10.1103/physrevb.81.214305},
  number = {21},
  pages = {214305},
  journal = {Physical Review B},
  publisher = {American Physical Society (APS)},
  author = {Sellan,  D. P. and Landry,  E. S. and Turney,  J. E. and McGaughey,  A. J. H. and Amon,  C. H.},
  year = {2010},
  month = jun 
}

@article{knoop_ab_2022,
  title = {Ab initio
 Green-Kubo simulations of heat transport in solids: Method and implementation},
  volume = {107},
  ISSN = {2469-9969},
  url = {http://dx.doi.org/10.1103/PhysRevB.107.224304},
  DOI = {10.1103/physrevb.107.224304},
  number = {22},
  pages = {224304},
  journal = {Physical Review B},
  publisher = {American Physical Society (APS)},
  author = {Knoop,  Florian and Scheffler,  Matthias and Carbogno,  Christian},
  year = {2023},
  month = jun 
}

@article{ercole_sportran_2022,
  title = {SporTran: A code to estimate transport coefficients from the cepstral analysis of (multivariate) current time series},
  volume = {280},
  ISSN = {0010-4655},
  url = {http://dx.doi.org/10.1016/j.cpc.2022.108470},
  DOI = {10.1016/j.cpc.2022.108470},
  journal = {Computer Physics Communications},
  publisher = {Elsevier BV},
  author = {Ercole,  Loris and Bertossa,  Riccardo and Bisacchi,  Sebastiano and Baroni,  Stefano},
  year = {2022},
  month = nov,
  pages = {108470}
}

@article{ercole_accurate_2017,
  title = {Accurate thermal conductivities from optimally short molecular dynamics simulations},
  volume = {7},
  ISSN = {2045-2322},
  url = {http://dx.doi.org/10.1038/s41598-017-15843-2},
  DOI = {10.1038/s41598-017-15843-2},
  number = {1},
  pages = {15835},
  journal = {Scientific Reports},
  publisher = {Springer Science and Business Media LLC},
  author = {Ercole,  Loris and Marcolongo,  Aris and Baroni,  Stefano},
  year = {2017},
  month = nov 
}

@article{pegolo_thermal_2024,
  title = {Thermal transport of glasses via machine learning driven simulations},
  volume = {11},
  ISSN = {2296-8016},
  url = {http://dx.doi.org/10.3389/fmats.2024.1369034},
  DOI = {10.3389/fmats.2024.1369034},
  pages = {1369034},
  journal = {Frontiers in Materials},
  publisher = {Frontiers Media SA},
  author = {Pegolo,  Paolo and Grasselli,  Federico},
  year = {2024},
  month = mar 
}

@article{malosso_viscosity_2022,
  title = {Viscosity in water from first-principles and deep-neural-network simulations},
  volume = {8},
  ISSN = {2057-3960},
  url = {http://dx.doi.org/10.1038/s41524-022-00830-7},
  DOI = {10.1038/s41524-022-00830-7},
  number = {1},
  pages = {139},
  journal = {npj Computational Materials},
  publisher = {Springer Science and Business Media LLC},
  author = {Malosso,  Cesare and Zhang,  Linfeng and Car,  Roberto and Baroni,  Stefano and Tisi,  Davide},
  year = {2022},
  month = jul 
}

@article{tisi_thermal_2024,
  title = {Thermal conductivity of 
${\mathrm{{Li}}}_{3}{\mathrm{{PS}}}_{4}$
 solid electrolytes with 
ab initio
 accuracy},
  volume = {8},
  ISSN = {2475-9953},
  url = {http://dx.doi.org/10.1103/PhysRevMaterials.8.065403},
  DOI = {10.1103/physrevmaterials.8.065403},
  number = {6},
  pages = {065403},
  journal = {Physical Review Materials},
  publisher = {American Physical Society (APS)},
  author = {Tisi,  Davide and Grasselli,  Federico and Gigli,  Lorenzo and Ceriotti,  Michele},
  year = {2024},
  month = jun 
}

@article{ercole_gauge_2016,
  title = {Gauge Invariance of Thermal Transport Coefficients},
  volume = {185},
  ISSN = {1573-7357},
  url = {http://dx.doi.org/10.1007/s10909-016-1617-6},
  DOI = {10.1007/s10909-016-1617-6},
  number = {1–2},
  journal = {Journal of Low Temperature Physics},
  publisher = {Springer Science and Business Media LLC},
  author = {Ercole,  Loris and Marcolongo,  Aris and Umari,  Paolo and Baroni,  Stefano},
  year = {2016},
  month = apr,
  pages = {79–86}
}

@article{otero-lema_kute_2025,
  title = {KUTE: Green–Kubo Uncertainty-Based Transport Coefficient Estimator},
  volume = {65},
  ISSN = {1549-960X},
  url = {http://dx.doi.org/10.1021/acs.jcim.4c02219},
  DOI = {10.1021/acs.jcim.4c02219},
  number = {7},
  journal = {Journal of Chemical Information and Modeling},
  publisher = {American Chemical Society (ACS)},
  author = {Otero-Lema,  Martín and Lois-Cuns,  Raúl and Boado,  Miguel A. and Montes-Campos,  Hadrián and Méndez-Morales,  Trinidad and Varela,  Luis M.},
  year = {2025},
  month = mar,
  pages = {3477–3487}
}

@article{Oliveira2017,
  title = {Method to manage integration error in the Green-Kubo method},
  volume = {95},
  ISSN = {2470-0053},
  url = {http://dx.doi.org/10.1103/PhysRevE.95.023308},
  DOI = {10.1103/physreve.95.023308},
  number = {2},
  pages = {023308},
  journal = {Physical Review E},
  publisher = {American Physical Society (APS)},
  author = {Oliveira,  Laura de Sousa and Greaney,  P. Alex},
  year = {2017},
  month = feb 
}

@inproceedings{batatia_mace_2022,
 author = {Batatia, Ilyes and Kovacs, David P and Simm, Gregor and Ortner, Christoph and Csanyi, Gabor},
 booktitle = {Advances in Neural Information Processing Systems},
 editor = {S. Koyejo and S. Mohamed and A. Agarwal and D. Belgrave and K. Cho and A. Oh},
 pages = {11423--11436},
 publisher = {Curran Associates, Inc.},
 title = {MACE: Higher Order Equivariant Message Passing Neural Networks for Fast and Accurate Force Fields},
 url = {https://proceedings.neurips.cc/paper_files/paper/2022/file/4a36c3c51af11ed9f34615b81edb5bbc-Paper-Conference.pdf},
 volume = {35},
 year = {2022}
}

@misc{batatia_foundation_2023,
  doi = {10.48550/ARXIV.2401.00096},
  url = {https://arxiv.org/abs/2401.00096},
  author = {Batatia,  Ilyes and Benner,  Philipp and Chiang,  Yuan and Elena,  Alin M. and Kovács,  Dávid P. and Riebesell,  Janosh and Advincula,  Xavier R. and Asta,  Mark and Avaylon,  Matthew and Baldwin,  William J. and Berger,  Fabian and Bernstein,  Noam and Bhowmik,  Arghya and Blau,  Samuel M. and Cărare,  Vlad and Darby,  James P. and De,  Sandip and Della Pia,  Flaviano and Deringer,  Volker L. and Elijošius,  Rokas and El-Machachi,  Zakariya and Falcioni,  Fabio and Fako,  Edvin and Ferrari,  Andrea C. and Genreith-Schriever,  Annalena and George,  Janine and Goodall,  Rhys E. A. and Grey,  Clare P. and Grigorev,  Petr and Han,  Shuang and Handley,  Will and Heenen,  Hendrik H. and Hermansson,  Kersti and Holm,  Christian and Jaafar,  Jad and Hofmann,  Stephan and Jakob,  Konstantin S. and Jung,  Hyunwook and Kapil,  Venkat and Kaplan,  Aaron D. and Karimitari,  Nima and Kermode,  James R. and Kroupa,  Namu and Kullgren,  Jolla and Kuner,  Matthew C. and Kuryla,  Domantas and Liepuoniute,  Guoda and Margraf,  Johannes T. and Magdău,  Ioan-Bogdan and Michaelides,  Angelos and Moore,  J. Harry and Naik,  Aakash A. and Niblett,  Samuel P. and Norwood,  Sam Walton and O'Neill,  Niamh and Ortner,  Christoph and Persson,  Kristin A. and Reuter,  Karsten and Rosen,  Andrew S. and Schaaf,  Lars L. and Schran,  Christoph and Shi,  Benjamin X. and Sivonxay,  Eric and Stenczel,  Tamás K. and Svahn,  Viktor and Sutton,  Christopher and Swinburne,  Thomas D. and Tilly,  Jules and van der Oord,  Cas and Varga-Umbrich,  Eszter and Vegge,  Tejs and Vondrák,  Martin and Wang,  Yangshuai and Witt,  William C. and Zills,  Fabian and Csányi,  Gábor},
  title = {A foundation model for atomistic materials chemistry},
  publisher = {arXiv},
  year = {2024},
  copyright = {Creative Commons Attribution Non Commercial No Derivatives 4.0 International}
}

@article{langer_heat_2023,
  title = {Heat flux for semilocal machine-learning potentials},
  volume = {108},
  ISSN = {2469-9969},
  url = {http://dx.doi.org/10.1103/PhysRevB.108.L100302},
  DOI = {10.1103/physrevb.108.l100302},
  number = {10},
  pages = {L100302},
  journal = {Physical Review B},
  publisher = {American Physical Society (APS)},
  author = {Langer,  Marcel F. and Knoop,  Florian and Carbogno,  Christian and Scheffler,  Matthias and Rupp,  Matthias},
  year = {2023},
  month = sep 
}

@article{langer_stress_2023,
  title = {Stress and heat flux via automatic differentiation},
  volume = {159},
  ISSN = {1089-7690},
  url = {http://dx.doi.org/10.1063/5.0155760},
  DOI = {10.1063/5.0155760},
  number = {17},
  pages = {174105},
  journal = {The Journal of Chemical Physics},
  publisher = {AIP Publishing},
  author = {Langer,  Marcel F. and Frank,  J. Thorben and Knoop,  Florian},
  year = {2023},
  month = nov 
}

@misc{reddi_convergence_2019,
  doi = {10.48550/ARXIV.1904.09237},
  url = {https://arxiv.org/abs/1904.09237},
  author = {Reddi,  Sashank J. and Kale,  Satyen and Kumar,  Sanjiv},
  title = {On the Convergence of Adam and Beyond},
  publisher = {arXiv},
  year = {2019},
  copyright = {arXiv.org perpetual,  non-exclusive license}
}

@article{heid_spatially_2024,
  title = {Spatially Resolved Uncertainties for Machine Learning Potentials},
  volume = {64},
  ISSN = {1549-960X},
  url = {http://dx.doi.org/10.1021/acs.jcim.4c00904},
  DOI = {10.1021/acs.jcim.4c00904},
  number = {16},
  journal = {Journal of Chemical Information and Modeling},
  publisher = {American Chemical Society (ACS)},
  author = {Heid,  Esther and Sch\"{o}rghuber,  Johannes and Wanzenb\"{o}ck,  Ralf and Madsen,  Georg K. H.},
  year = {2024},
  month = aug,
  pages = {6377–6387}
}

@article{Carrete2023,
  title = {Deep ensembles vs committees for uncertainty estimation in neural-network force fields: Comparison and application to active learning},
  volume = {158},
  ISSN = {1089-7690},
  url = {http://dx.doi.org/10.1063/5.0146905},
  DOI = {10.1063/5.0146905},
  number = {20},
  pages = {204801},
  journal = {The Journal of Chemical Physics},
  publisher = {AIP Publishing},
  author = {Carrete,  Jesús and Montes-Campos,  Hadrián and Wanzenb\"{o}ck,  Ralf and Heid,  Esther and Madsen,  Georg K. H.},
  year = {2023},
  month = may 
}

@article{schwalbe-koda_differentiable_2021,
  title = {Differentiable sampling of molecular geometries with uncertainty-based adversarial attacks},
  volume = {12},
  ISSN = {2041-1723},
  url = {http://dx.doi.org/10.1038/s41467-021-25342-8},
  DOI = {10.1038/s41467-021-25342-8},
  number = {1},
  pages = {5104},
  journal = {Nature Communications},
  publisher = {Springer Science and Business Media LLC},
  author = {Schwalbe-Koda,  Daniel and Tan,  Aik Rui and Gómez-Bombarelli,  Rafael},
  year = {2021},
  month = aug 
}

@article{Axelrod2022,
  title = {Learning Matter: Materials Design with Machine Learning and Atomistic Simulations},
  volume = {3},
  ISSN = {2643-6728},
  url = {http://dx.doi.org/10.1021/accountsmr.1c00238},
  DOI = {10.1021/accountsmr.1c00238},
  number = {3},
  journal = {Accounts of Materials Research},
  publisher = {American Chemical Society (ACS)},
  author = {Axelrod,  Simon and Schwalbe-Koda,  Daniel and Mohapatra,  Somesh and Damewood,  James and Greenman,  Kevin P. and Gómez-Bombarelli,  Rafael},
  year = {2022},
  month = feb,
  pages = {343–357}
}

@article{pan_machine_2024,
  title = {Machine learning boosted ab initio
 study of the thermal conductivity of Janus PtSTe van der Waals heterostructures},
  volume = {109},
  ISSN = {2469-9969},
  url = {http://dx.doi.org/10.1103/PhysRevB.109.035417},
  DOI = {10.1103/physrevb.109.035417},
  number = {3},
  pages = {035417},
  journal = {Physical Review B},
  publisher = {American Physical Society (APS)},
  author = {Pan,  Lijun and Carrete,  Jesús and Wang,  Zhao and Madsen,  Georg K. H.},
  year = {2024},
  month = jan 
}

@article{schorghuber_flat_2025,
  title = {From flat to stepped: active learning frameworks for investigating local structure at copper–water interfaces},
  volume = {27},
  ISSN = {1463-9084},
  url = {http://dx.doi.org/10.1039/D5CP00396B},
  DOI = {10.1039/d5cp00396b},
  number = {17},
  journal = {Physical Chemistry Chemical Physics},
  publisher = {Royal Society of Chemistry (RSC)},
  author = {Sch\"{o}rghuber,  Johannes and Bučková,  Nina and Heid,  Esther and Madsen,  Georg K. H.},
  year = {2025},
  pages = {9169–9177}
}

@article{Togo2015,
  title = {First principles phonon calculations in materials science},
  volume = {108},
  ISSN = {1359-6462},
  url = {http://dx.doi.org/10.1016/j.scriptamat.2015.07.021},
  DOI = {10.1016/j.scriptamat.2015.07.021},
  journal = {Scripta Materialia},
  publisher = {Elsevier BV},
  author = {Togo,  Atsushi and Tanaka,  Isao},
  year = {2015},
  month = nov,
  pages = {1–5}
}

@article{Kresse1993,
  title = {Ab initiomolecular dynamics for liquid metals},
  volume = {47},
  ISSN = {1095-3795},
  url = {http://dx.doi.org/10.1103/PhysRevB.47.558},
  DOI = {10.1103/physrevb.47.558},
  number = {1},
  journal = {Physical Review B},
  publisher = {American Physical Society (APS)},
  author = {Kresse,  G. and Hafner,  J.},
  year = {1993},
  month = jan,
  pages = {558–561}
}

@article{Kresse1994,
  title = {Ab initiomolecular-dynamics simulation of the liquid-metal–amorphous-semiconductor transition in germanium},
  volume = {49},
  ISSN = {1095-3795},
  url = {http://dx.doi.org/10.1103/PhysRevB.49.14251},
  DOI = {10.1103/physrevb.49.14251},
  number = {20},
  journal = {Physical Review B},
  publisher = {American Physical Society (APS)},
  author = {Kresse,  G. and Hafner,  J.},
  year = {1994},
  month = may,
  pages = {14251–14269}
}

@article{Kresse1996,
	title = {Efficiency of ab-initio total energy calculations for metals and semiconductors using a plane-wave basis set},
	volume = {6},
	issn = {09270256},
	doi = {10.1016/0927-0256(96)00008-0},
	number = {1},
	journal = {Computational Materials Science},
	author = {Kresse, G. and Furthmüller, J.},
	year = {1996},
	pages = {15--50},
}

@article{Kresse1996_2,
	title = {Efficient iterative schemes for ab initio total-energy calculations using a plane-wave basis set},
	volume = {54},
	issn = {1550235X},
	doi = {10.1103/PhysRevB.54.11169},
	number = {16},
	journal = {Physical Review B - Condensed Matter and Materials Physics},
	author = {Kresse, G. and Furthmüller, J.},
	year = {1996},
	pages = {11169--11186},
}

@article{blochl_projector_1994,
	title = {Projector augmented-wave method},
	volume = {50},
	issn = {01631829},
	doi = {10.1103/PhysRevB.50.17953},
	number = {24},
	journal = {Physical Review B},
	author = {Blöchl, P. E.},
	year = {1994},
	pmid = {9976227},
	pages = {17953--17979},
}

@article{Perdew1996,
	title = {Generalized gradient approximation made simple},
	volume = {77},
	issn = {10797114},
	doi = {10.1103/PhysRevLett.77.3865},
	number = {18},
	journal = {Physical Review Letters},
	author = {Perdew, John P. and Burke, Kieron and Ernzerhof, Matthias},
	year = {1996},
	pmid = {10062328},
	pages = {3865--3868},
}

@article{perdew_restoring_2008,
	title = {Restoring the {Density}-{Gradient} {Expansion} for {Exchange} in {Solids} and {Surfaces}},
	volume = {100},
    number = {13},
    pages = {136406},
	journal = {Physical Review Letters},
	author = {Perdew, John P. and Ruzsinszky, Adrienn and Csonka, Gábor I. and Vydrov, Oleg A. and Scuseria, Gustavo E. and Constantin, Lucian A. and Zhou, Xiaolan and Burke, Kieron},
	year = {2008},
}

@article{perdew_generalized_1997,
	title = {Generalized {Gradient} {Approximation} {Made} {Simple} [{Phys}. {Rev}. {Lett}. 77, 3865 (1996)]},
	volume = {78},
	issn = {0031-9007},
	url = {http://www.ncbi.nlm.nih.gov/pubmed/22502509%5Cnhttp://www.ncbi.nlm.nih.gov/pubmed/10062328%5Cnhttp://link.aps.org/doi/10.1103/PhysRevLett.77.3865},
	number = {7},
	journal = {Physical Review Letters},
	author = {Perdew, John P. and Burke, Kieron and Ernzerhof, Matthias},
	year = {1997},
	pmid = {10062328},
	note = {ISBN: 9780596529321},
	pages = {1396--1396},
}

@article{Kubo1966,
	title = {The fluctuation–dissipation theorem},
	volume = {29},
	issn = {13665812},
	doi = {10.1080/00107514.2017.1298289},
	number = {1},
	journal = {Reports on Progress in Physics},
	author = {Kubo, R.},
	year = {1966},
	pages = {255--284},
}

@article{hardy_energy-flux_1968,
	title = {Energy-{Flux} {Operator} for a {Lattice}},
	volume = {132},
	journal = {Physical Review},
	author = {Hardy, Robert J},
    number = {1},
    pages={168},
    publisher = {APS},
	year = {1968},
}

@article{Hamakawa2025,
  title = {Accurate heat flux formula and thermal conductivity calculation in molecular dynamics simulations with machine learning potentials},
  volume = {138},
  ISSN = {1089-7550},
  url = {http://dx.doi.org/10.1063/5.0278501},
  DOI = {10.1063/5.0278501},
  number = {5},
  pages = {055105},
  journal = {Journal of Applied Physics},
  publisher = {AIP Publishing},
  author = {Hamakawa,  Tomu and McGaughey,  Alan J. H. and Shiomi,  Junichiro},
  year = {2025},
  month = aug 
}

@article{Thompson2022,
	title = {{LAMMPS} - a flexible simulation tool for particle-based materials modeling at the atomic, meso, and continuum scales},
	volume = {271},
	issn = {00104655},
	url = {https://doi.org/10.1016/j.cpc.2021.108171},
	doi = {10.1016/j.cpc.2021.108171},
	journal = {Computer Physics Communications},
	author = {Thompson, Aidan P. and Aktulga, H. Metin and Berger, Richard and Bolintineanu, Dan S. and Brown, W. Michael and Crozier, Paul S. and in 't Veld, Pieter J. and Kohlmeyer, Axel and Moore, Stan G. and Nguyen, Trung Dac and Shan, Ray and Stevens, Mark J. and Tranchida, Julien and Trott, Christian and Plimpton, Steven J.},
	year = {2022},
	note = {Publisher: Elsevier B.V.},
	pages = {108171},
}

@article{hjorth_larsen_atomic_2017,
	title = {The atomic simulation environment—a {Python} library for working with atoms},
	volume = {29},
	issn = {0953-8984},
	url = {https://dx.doi.org/10.1088/1361-648X/aa680e},
	doi = {10.1088/1361-648X/aa680e},
	language = {en},
	number = {27},
	urldate = {2025-02-17},
	journal = {Journal of Physics: Condensed Matter},
	author = {Hjorth Larsen, Ask and Jørgen Mortensen, Jens and Blomqvist, Jakob and Castelli, Ivano E and Christensen, Rune and Dułak, Marcin and Friis, Jesper and Groves, Michael N and Hammer, Bjørk and Hargus, Cory and Hermes, Eric D and Jennings, Paul C and Bjerre Jensen, Peter and Kermode, James and Kitchin, John R and Leonhard Kolsbjerg, Esben and Kubal, Joseph and Kaasbjerg, Kristen and Lysgaard, Steen and Bergmann Maronsson, Jón and Maxson, Tristan and Olsen, Thomas and Pastewka, Lars and Peterson, Andrew and Rostgaard, Carsten and Schiøtz, Jakob and Schütt, Ole and Strange, Mikkel and Thygesen, Kristian S and Vegge, Tejs and Vilhelmsen, Lasse and Walter, Michael and Zeng, Zhenhua and Jacobsen, Karsten W},
	month = jun,
	year = {2017},
	note = {Publisher: IOP Publishing},
	pages = {273002},
}

@article{burnham_multimodel_2004,
	title = {Multimodel {Inference}: {Understanding} {AIC} and {BIC} in {Model} {Selection}},
	volume = {33},
	issn = {0049-1241},
	shorttitle = {Multimodel {Inference}},
	url = {https://doi.org/10.1177/0049124104268644},
	doi = {10.1177/0049124104268644},
	language = {en},
	number = {2},
	urldate = {2024-10-31},
	journal = {Sociological Methods \& Research},
	author = {Burnham, Kenneth P. and Anderson, David R.},
	month = nov,
	year = {2004},
	note = {Publisher: SAGE Publications Inc},
	pages = {261--304},
}

@article{lukacs_model_2010,
	title = {Model selection bias and {Freedman}’s paradox},
	volume = {62},
	issn = {1572-9052},
	url = {https://doi.org/10.1007/s10463-009-0234-4},
	doi = {10.1007/s10463-009-0234-4},
	language = {en},
	number = {1},
	urldate = {2024-11-08},
	journal = {Annals of the Institute of Statistical Mathematics},
	author = {Lukacs, Paul M. and Burnham, Kenneth P. and Anderson, David R.},
	month = feb,
	year = {2010},
	pages = {117--125},
}

@misc{wieser_supplementary_2025,
	title = {Supplementary data for "{Accelerating} first-principles molecular-dynamics thermal conductivity calculations for complex systems"},
	url = {https://zenodo.org/records/16913162},
	doi = {10.5281/zenodo.16913162},
	urldate = {2025-08-21},
	publisher = {Zenodo},
	author = {Wieser, Sandro and Cen, Yu-Jie and Madsen, Georg Kent Hellerup and Carrete, Jesús},
	month = aug,
	year = {2025},
}

@misc{mace-unfolded_2025,
	title = {mace-unfolded},
    author = {Wieser, Sandro},
	copyright = {Apache-2.0},
	url = {https://github.com/pulgon-project/mace-unfolded},
	urldate = {2025-08-21},
	publisher = {PULGON},
	month = jul,
	year = {2025},
	note = {(accessed 2025-09-05)},
}

@article{Zhou2017,
  title = {Nonmonotonic Diameter Dependence of Thermal Conductivity of Extremely Thin Si Nanowires: Competition between Hydrodynamic Phonon Flow and Boundary Scattering},
  volume = {17},
  ISSN = {1530-6992},
  url = {http://dx.doi.org/10.1021/acs.nanolett.6b05113},
  DOI = {10.1021/acs.nanolett.6b05113},
  number = {2},
  journal = {Nano Letters},
  publisher = {American Chemical Society (ACS)},
  author = {Zhou,  Yanguang and Zhang,  Xiaoliang and Hu,  Ming},
  year = {2017},
  month = jan,
  pages = {1269–1276}
}

@article{doi:10.1021/acs.jctc.9b01174,
author = {Marcolongo, Aris and Ercole, Loris and Baroni, Stefano},
title = {Gauge Fixing for Heat-Transport Simulations},
journal = {J. Chem. Theory Comput.},
volume = {16},
number = {5},
pages = {3352-3362},
year = {2020},
doi = {10.1021/acs.jctc.9b01174},
URL = {https://doi.org/10.1021/acs.jctc.9b01174},
}

@Misc{JCGMGUM,
  Title                    = {Evaluation of measurement data --- {G}uide to the expression of uncertainty in measurement},

  Author                   = {BIPM and IEC and IFCC and ILAC and ISO and IUPAC and IUPAP and OIML},
  HowPublished             = {Joint Committee for Guides in Metrology, JCGM 100:2008},
  doi                      = {https://doi.org/10.59161/JCGM100-2008E}
}

@misc{Hendrycks2016,
  doi = {10.48550/ARXIV.1606.08415},
  url = {https://arxiv.org/abs/1606.08415},
  author = {Hendrycks,  Dan and Gimpel,  Kevin},
  title = {Gaussian Error Linear Units (GELUs)},
  publisher = {arXiv},
  year = {2016},
  copyright = {arXiv.org perpetual,  non-exclusive license}
}

\end{document}



\begin{figure}
    \centering
    \includegraphics[width=\linewidth]{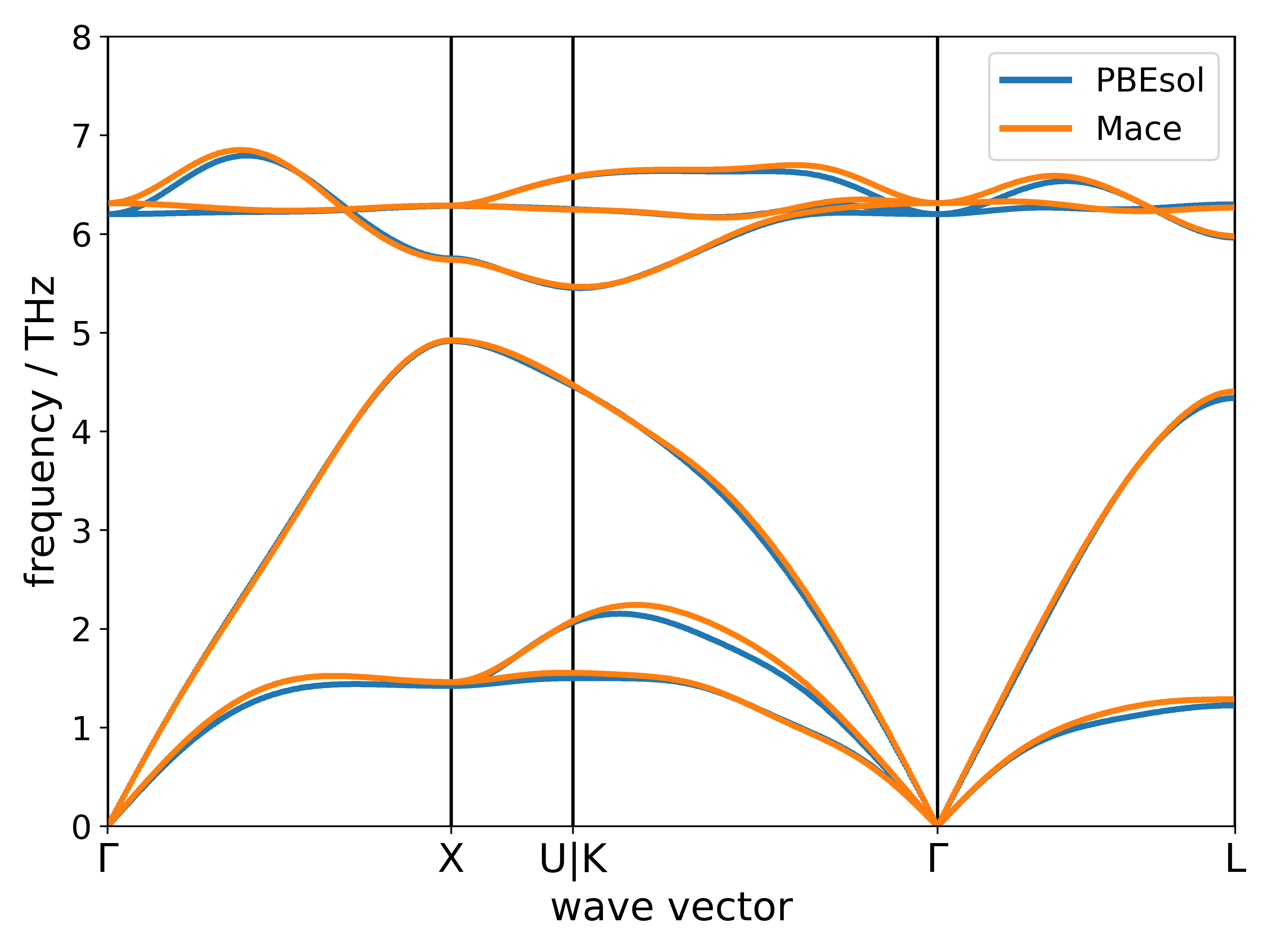}
    \caption{Phonon band structure of the InAs ZB bulk phase computed either with DFT (PBEsol) or using the trained MACE model. }
    \label{f_ZB_band_comparison}
\end{figure}


\begin{figure}
    \centering
    \includegraphics[width=\linewidth]{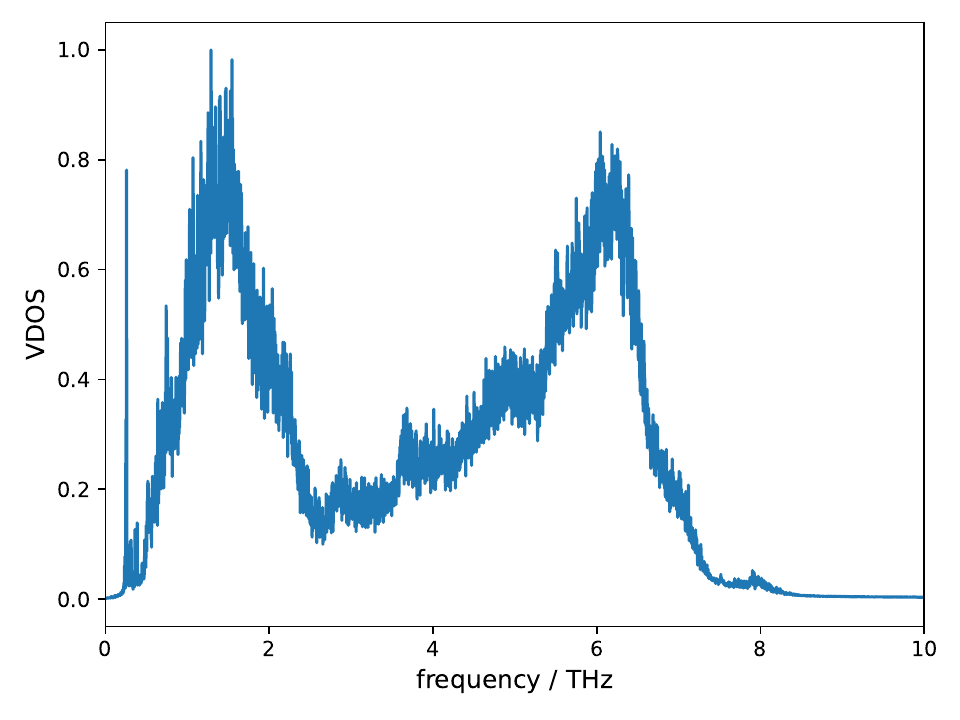}
    \caption{Vibrational density of states as obtained from the velocity autocorrelation function from the first \SI{100}{\pico\second} of a molecular dynamics simulation for the zincblende nanowire.}
    \label{f_vdos_ZB}
\end{figure}

\begin{figure}
    \centering
    \includegraphics[width=\linewidth]{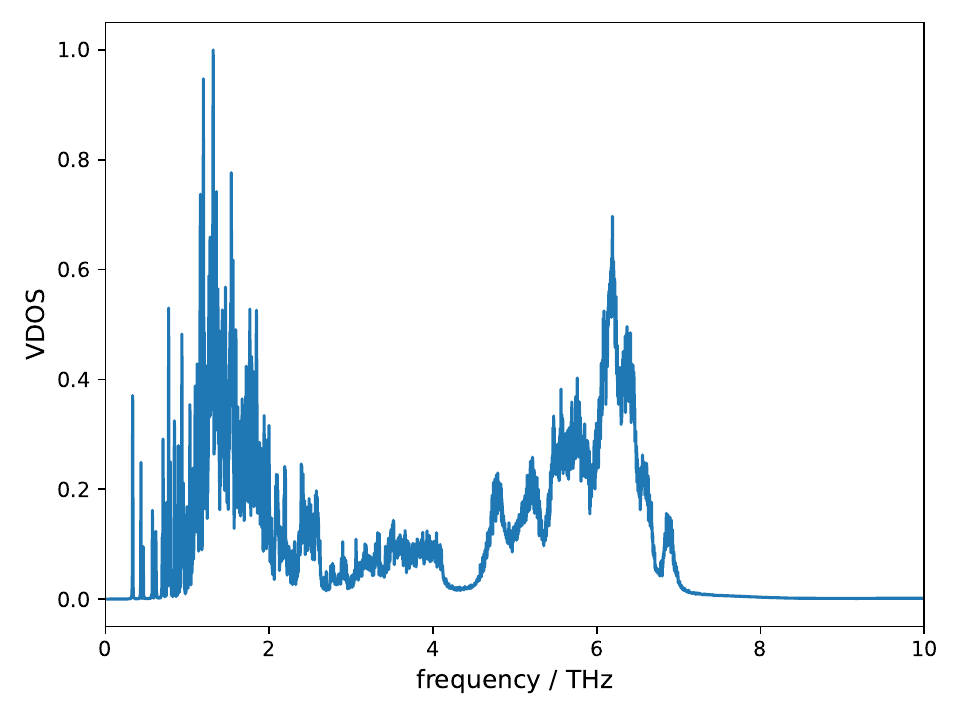}
    \caption{Vibrational density of states as obtained from the velocity autocorrelation function from the first \SI{100}{\pico\second} of a molecular dynamics simulation for the wurtzite nanowire.}
    \label{f_vdos_WZ}
\end{figure}

\begin{figure}
    \centering
    \includegraphics[width=0.7\linewidth]{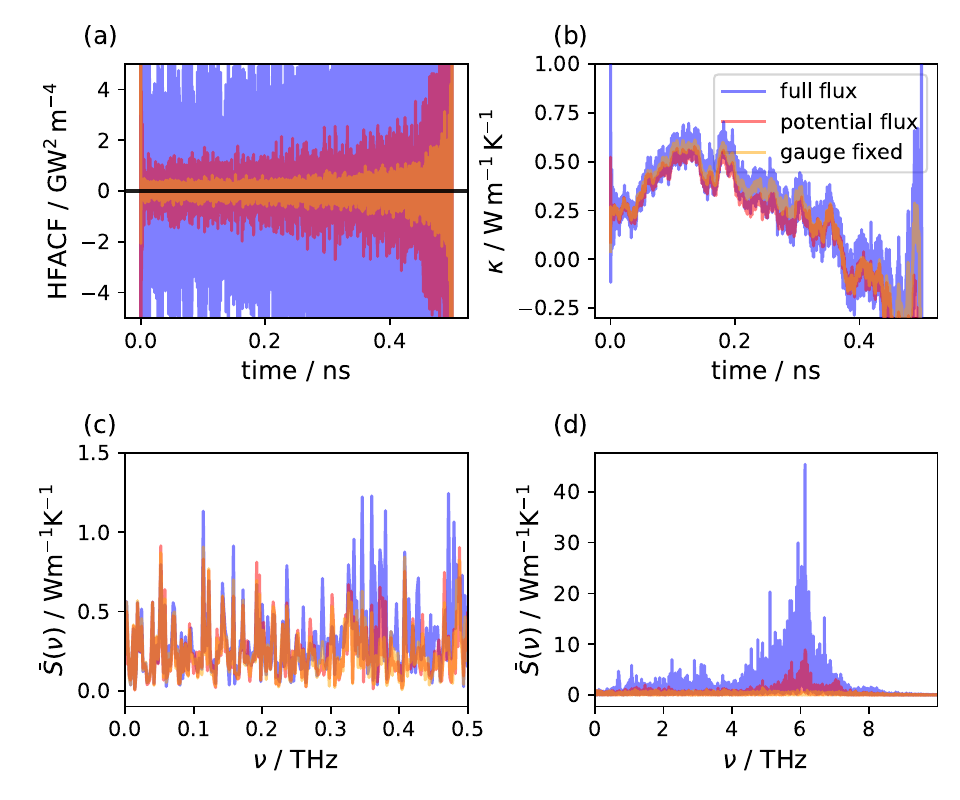}
    \caption{Comparison of different noise reduction techniques for the independent simulation number 1 for the ZB nanowire: using the full flux, removing the convective term leaving only the potential flux, and exploiting the gauge invariance of the flux. Comparisons shown are of (a) the heat flux autocorrelation functions, (b) the thermal conductivity $\kappa$, (c) the low-frequency power spectrum, and (d) the full power spectrum.}
    \label{f_spectral_discussion_0}
\end{figure}

\newcounter{x}
\forloop{x}{1}{\value{x}<11}{
  \begin{figure}[ht]
    \centering
    \includegraphics[width=0.7\linewidth]{images/methods/GK/individual_spectra/acf_22A-zb-hex_\arabic{x}.pdf}
    \caption{Comparison of different noise reduction techniques for the independent simulation number 
      \number\numexpr\value{x}+1\relax\ for the ZB nanowire in the same style as figure~\ref{f_spectral_discussion_0}.}
    \label{f_spectral_discussion_\arabic{x}}
  \end{figure}
}

\begin{figure}
    \centering
    \includegraphics[width=\linewidth]{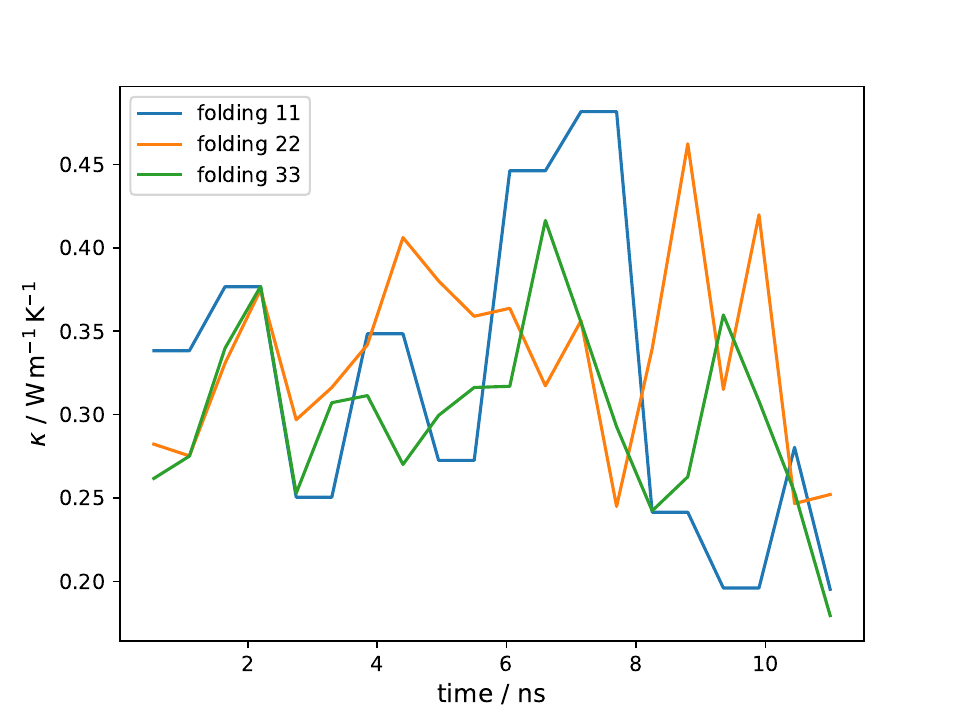}
    \caption{Figure showcasing the zero frequency thermal conductivity from the power spectrum for the ZB nanowire.}
    \label{f_zero_freq_plot_ZB}
\end{figure}

\begin{figure}
    \centering
    \includegraphics[width=\linewidth]{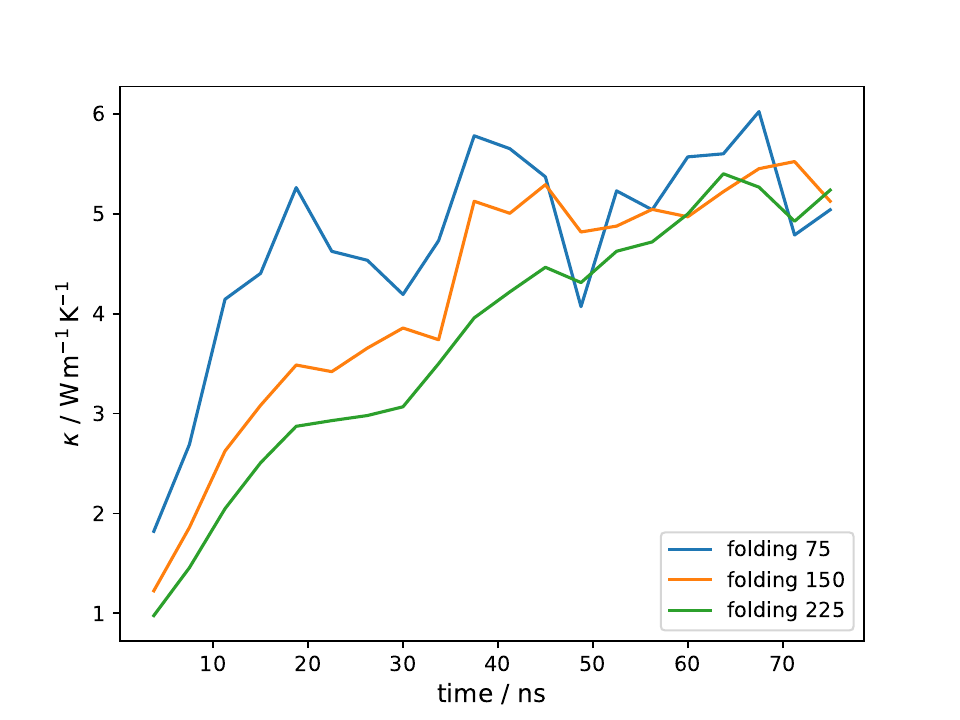}
    \caption{Figure showcasing the zero frequency thermal conductivity from the power spectrum for the WZ nanowire.}
    \label{f_zero_freq_plot_WZ}
\end{figure}

\begin{figure}
    \centering
    \includegraphics[width=0.9\linewidth]{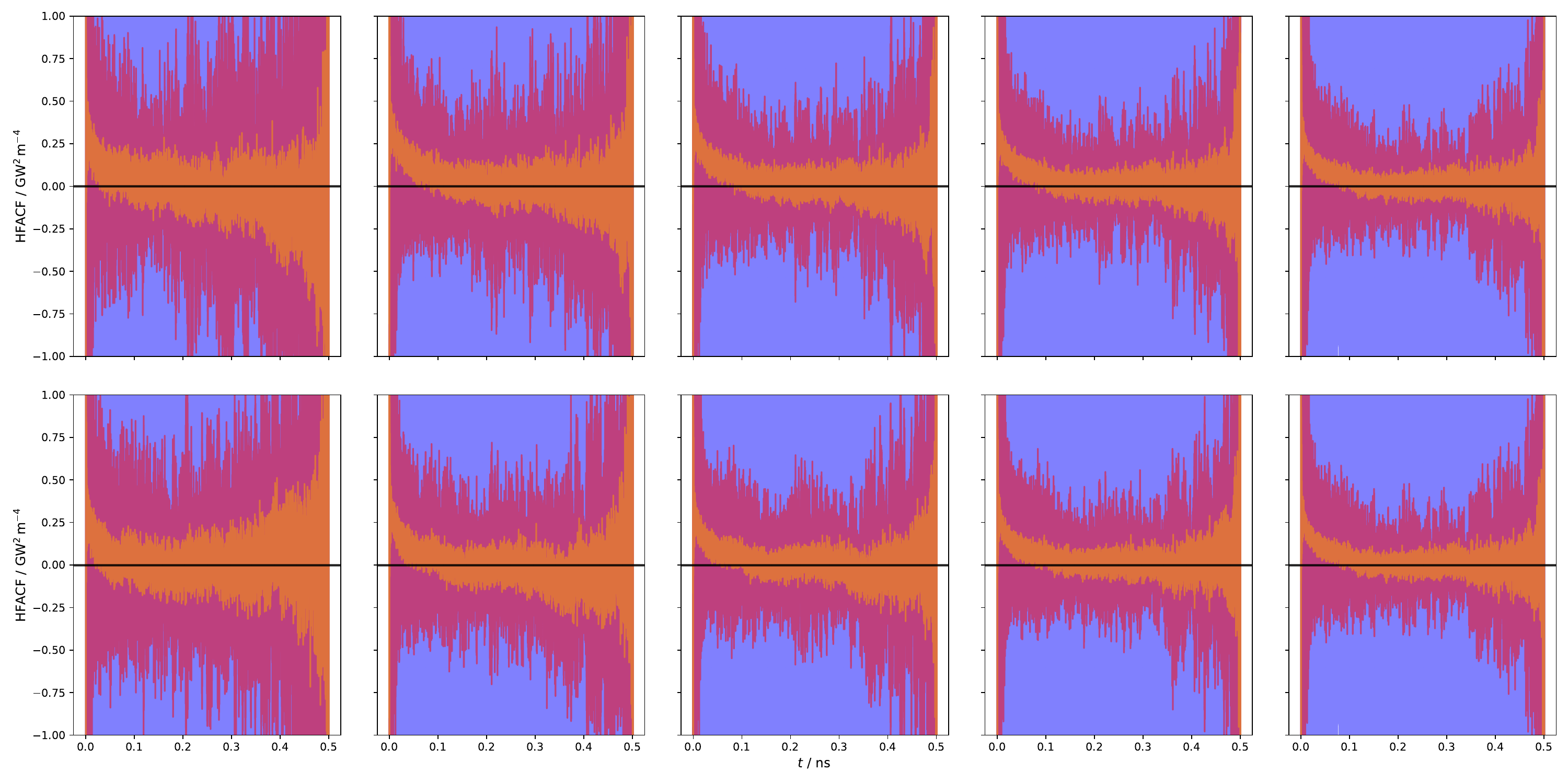}
    \caption{HFACFs depending on wether the number of independent simulations is increased (first row) or the total simulation time for each independent simulations (second row). The data is based on 5 \SI{5}{\nano\second} long MD trajectories for the WZ nanowire. In the far left figures \SI{5}{\nano\second} of total simulation time is used while on the far right figure it accumulated to \SI{25}{\nano\second}}
    \label{f_noise_choice}
\end{figure}

\begin{figure}
    \centering
    \includegraphics[width=0.9\linewidth]{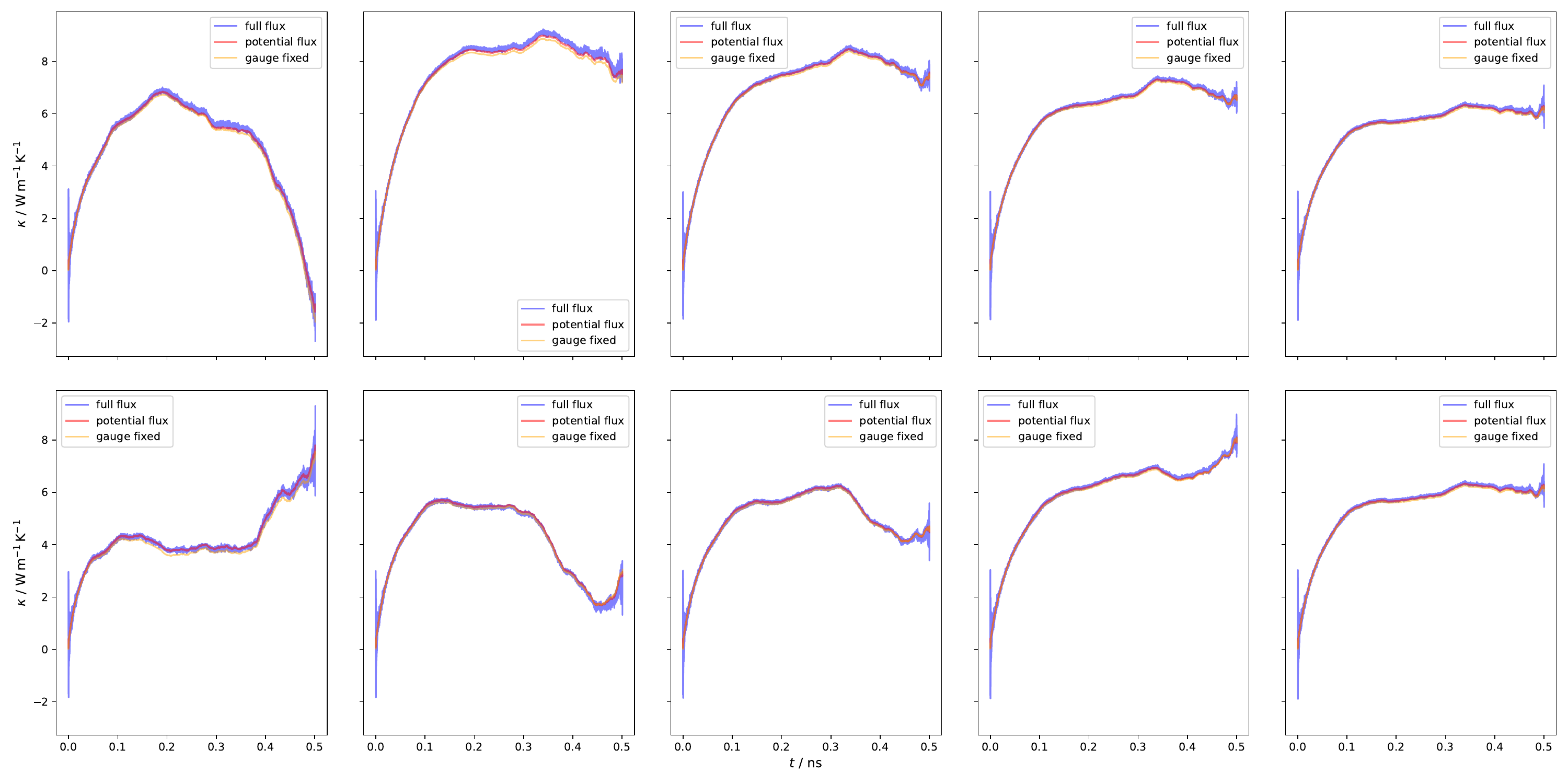}
    \caption{Thermal integral over the HFACFs depending on wether the number of independent simulations is increased (first row) or the total simulation time for each independent simulations (second row). The underlying data is the same as in Figure~\ref{f_noise_choice}}
    \label{f_noise_choice_kappa}
\end{figure}

\begin{figure}
    \centering
    \includegraphics[width=\linewidth]{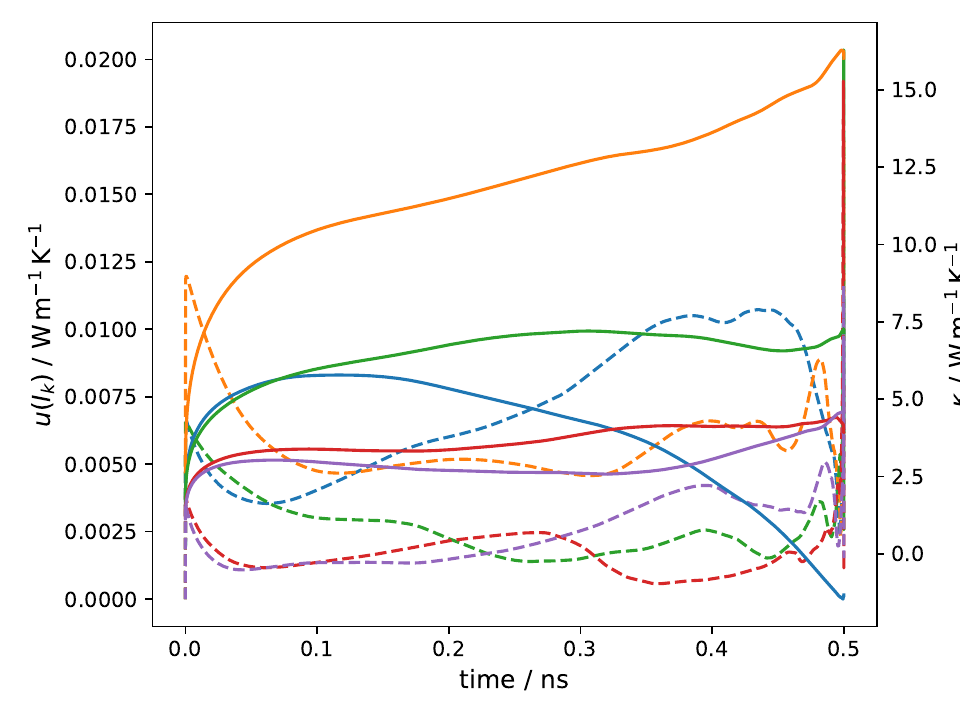}
    \caption{Thermal conductivities (solid lines) and uncertainties (dashed lines) from 5 individual 5 ns simulations for the WZ nanowire according to the prescribed KUTE approach after applying weighted averages to the individual runs.}
    \label{f_uncertainty_kute_individual}
\end{figure}

\begin{figure}
    \centering
    \includegraphics[width=\linewidth]{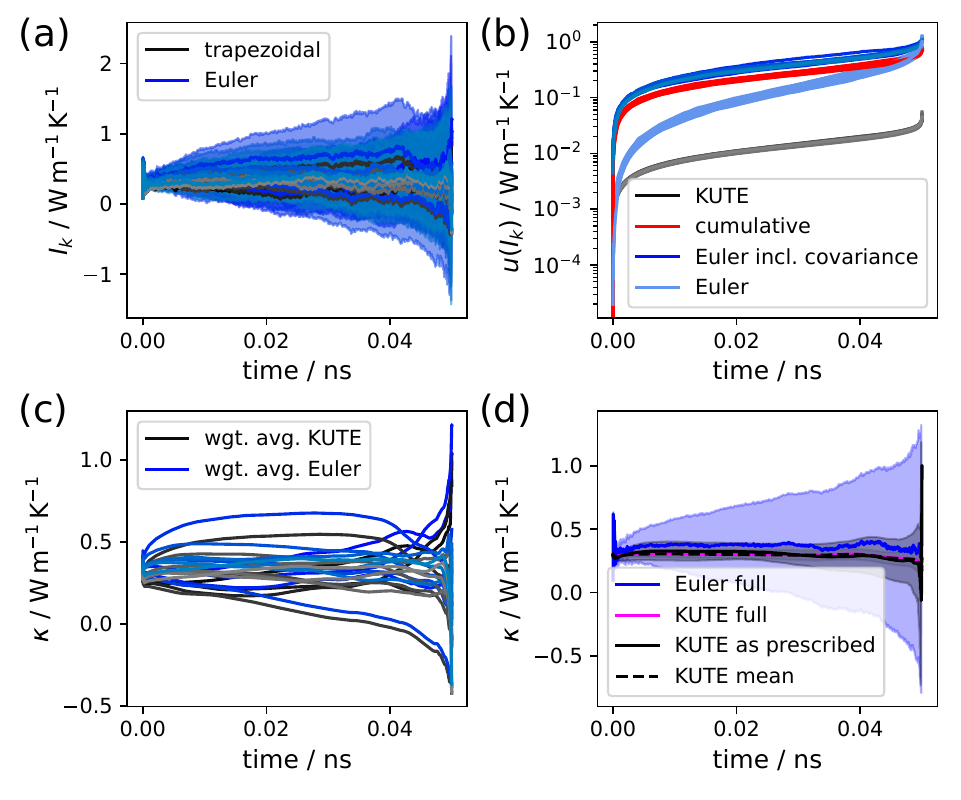}
    \caption{Comparison of analysis methods based on 11 1 ns GK simulations of the ZB nanowire. (a) Direct integration of a 20 times folded HFACF for the 11 independent runs using trapezoidal integration (black) or Euler integration (blue). The shaded area represents the uncertainty. (b) comparison of different uncertainty metrics. (c) Uncertainty weighted averages using the uncertainty including the covariance (blue) or the uncertainty as prescribed in the KUTE approach (black). (d) Comparison of different analysis techniques of the thermal conductivity with their uncertainty prediction. For a detailed description, see the main manuscript.}
    \label{f_uncertainty_discussion_ZB}
\end{figure}

\begin{figure}
    \centering
    \includegraphics[width=\linewidth]{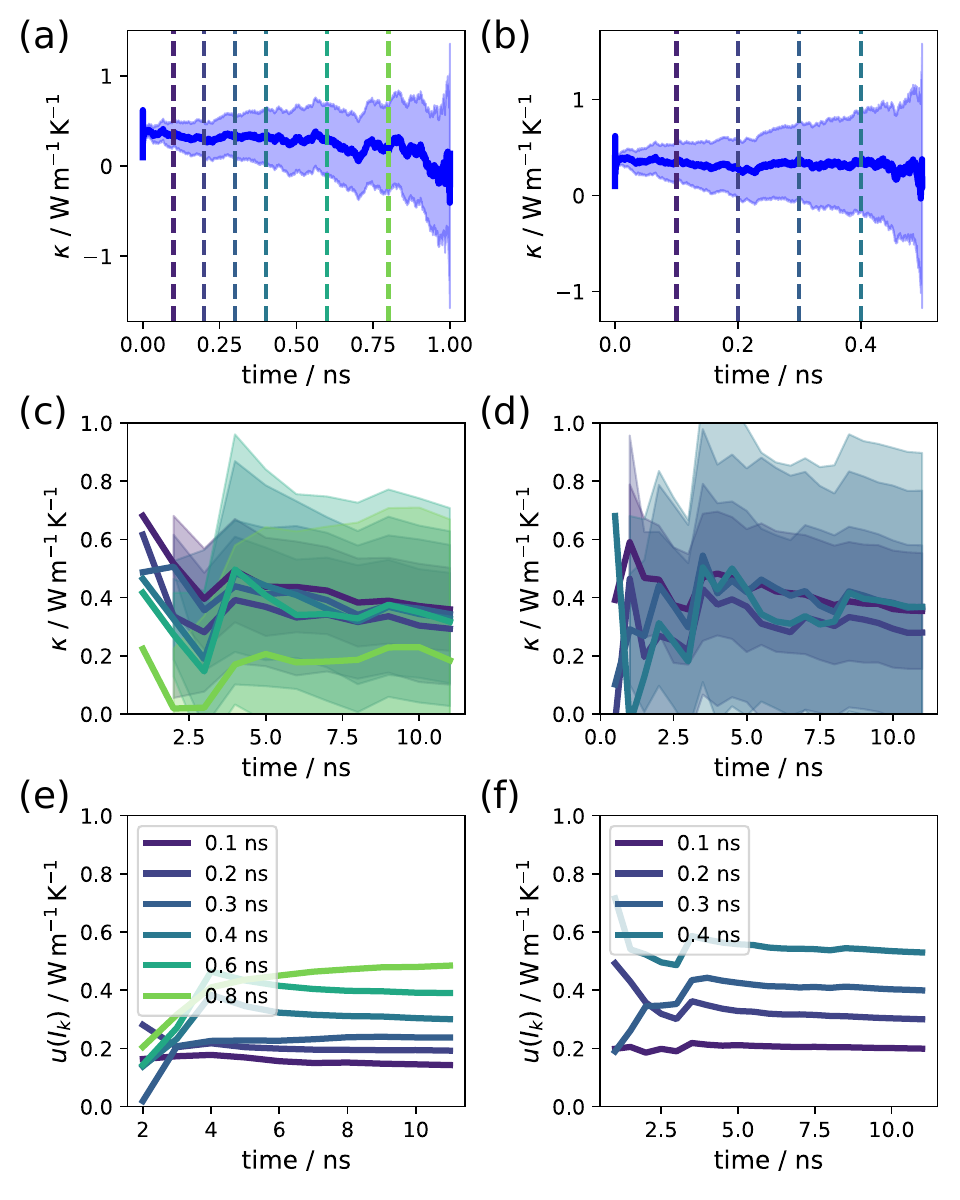}
    \caption{Time convergence of the ZB nanowire with a correlation time of 1 ns (a, c, e) or 0.5 ns (b, d, f) evaluated over a total simulation time of 11 ns. (a, b) The cumulative integrals performed using the Euler method and the corresponding uncertainty for the entire duration. (c, d) The time convergence curves based on values extracted after specific correlation times as indicated in (a,b). The corresponding uncertainties are shown in (e,f). }
    \label{f_euler_convergence_ZB}
\end{figure}